\documentclass[%
twocolumn,
aps,
prx,
amsmath,amssymb,
superscriptaddress,
]{revtex4-2}

\usepackage{graphicx}
\graphicspath{{./svg-inkscape}{./Figures/}}

\usepackage{dcolumn}
\usepackage{bm}
\usepackage[mathlines]{lineno}
\usepackage[utf8]{inputenc}
\usepackage[T1]{fontenc}
\usepackage{mathptmx}
\usepackage[colorlinks=true, linkcolor=black, citecolor=black, urlcolor=blue]{hyperref}

\usepackage{physics, amsmath, nicefrac}
\usepackage{ulem} 
\usepackage{xcolor}
\usepackage{svg} 
\usepackage{array, tabularx} 

\newcommand{\units}[1]{\,{\rm #1}}
\newcommand{\cell}[1]{$U_\text{#1}$} 
\newcommand{\dcell}[1]{$\vec{U}_\text{#1}$} 

\begin{document}

\title{Eliminating radiative losses in long-range exciton transport}

\author{Scott Davidson}
\email{sd109@hw.ac.uk}
\affiliation{SUPA, Institute of Photonics and Quantum Sciences, Heriot-Watt University, EH14 4AS, United Kingdom}

\author{Felix A. Pollock}
\affiliation{School of Physics and Astronomy, Monash University, Clayton, Victoria 3800, Australia}

\author{Erik Gauger}
\affiliation{SUPA, Institute of Photonics and Quantum Sciences, Heriot-Watt University, EH14 4AS, United Kingdom}

\begin{abstract}
    We demonstrate that it is possible to effectively eliminate radiative losses during excitonic energy transport in systems with an intrinsic energy gradient. By considering chain-like systems of repeating `unit' cells which can each consist of multiple sites, we show that tuning a single system parameter (the intra-unit-cell coupling) leads to efficient and highly robust transport over relatively long distances. This remarkable transport performance is shown to originate from a partitioning of the system's eigenstates into energetically-separated bright and dark subspaces, allowing long range transport to proceed efficiently through a `dark chain' of eigenstates. Finally, we discuss the effects of intrinsic dipole moments, which are of particular relevance to molecular architectures, and demonstrate that appropriately-aligned dipoles can lead to additional protection against other (non-radiative) loss processes. Our dimensionless open quantum systems model is designed to be broadly applicable to a range of experimental platforms.
\end{abstract}

\maketitle

\section{Introduction}
\label{sec:intro}

Energy transport processes are ubiquitous in physics, and are vitally important to a variety of technological applications and life-supporting biochemical reactions. Of particular interest are the subset of energy transport processes which occur at the nanoscale; where the time and length scales involved straddle the boundary between our best classical and quantum mechanical descriptions of nature. These include the early stages of natural photosynthesis~\cite{Intro:PhotosythesisMech, Intro:PhotoExcitons}, where solar photons are captured, transported, then stored as chemical energy; as well as artificial photovoltaic devices~\cite{Intro:SolarCells}.

There has been much debate surrounding the extent to which natural photosynthetic systems may have evolved in order to harness quantum coherence (of various types) as a means to improve efficiency~\cite{Intro:Non-quantum-Photosyntesis,Intro:Vibronic-coherence-only,Intro:Short-coherence-timescales, Intro:quantum-bio-revisited-2020, QT:KassalCoherentClassification, QT:PhotosynthCoherentSupertransfer}; but, in any case, clear advantages of utilizing quantum effects have been predicted in artificial devices~\cite{OQS:ArtificialCoherentEnhancementReview-Nature, QT:Quantum-photocell-beats-SQ-limit-Scully, QT:QHE-Dorfman-Scully, QT:delocalization-enhances-photocell-Kais, QT:Strong-EM-coupling-transport-2, QT:LongRangeHoppingDET, QT:Scholes-coherence-protecting-Hamiltonians, OQS:Cao-cavity-mediated-transport, OQS:Dubail-arrow-head-Hamiltonians}. These results provide ample motivation for further investigation into function and efficiency of nanoscale transport processes at a level that includes quantum mechanical interference.

In many such processes, excitons (electron-hole pair quasi-particles) are the primary energy carriers. However, exciton states tend to be inherently unstable due the possibility of electron-hole recombination processes, which destroy the exciton and result in energy loss. In many organic photovoltaic devices, these exciton recombination processes are the primary bottleneck for device efficiency, via their effects on exciton diffusion lengths~\cite{QT:OPV-review, QT:OPV-nr-loss-1, QT:perovskite-solar-cell-nr-loss}. The two primary loss channels involved are \textit{non-radiative} recombination, where energy is dissipated as heat (e.g.~via emission of multiple phonons)~\cite{QT:Multi-phonon-decay}, and \textit{radiative} recombination, where the energy is lost via the emission of optical photons.

One particularly interesting quantum mechanical effect of relevance to quantum transport, and radiative recombination processes in particular, is collective light-matter coupling~\cite{OQS:orig-superradiance-Dicke, OQS:Dicke-model-review, Intro:collective-emission-IOP-book, QT:OptimalGeomSupertransferLH}; where wavefunction interference effects can lead to a non-uniform distribution of radiative loss rates for the delocalized excitonic eigenstates of the transport system, with certain states becoming more susceptible to radiative losses (so called `bright' states), and others exhibiting a reduction in radiative loss rate (`dark' states). Significant research interest has been devoted to the latter set of states, with a variety of proposals aiming to utilize these dark states in order to mitigate radiative losses and therefore improve transport performance.
A number of these works have focused on dark state photocells in small systems~\cite{QT:bio-inspired-dark-state-Creatore, QT:Amir-photocell-prl, QT:chain-dark-state-photocell}, often assuming degenerate on-site energies throughout. Others have considered the effects of strong system-phonon couplings~\cite{QT:Rouse-strong-dark-state-dimers} or the competition between dark state protection and exciton delocalization~\cite{QT:dark-state-trapping-tradeoff}. More recently, larger systems have been investigated in the context of incoherent exciton diffusion across chains of biological light-harvesting complexes (with a flat on-site energy landscape)~\cite{QT:Plenio-PRX} and novel energetic landscapes have been shown to significantly enhance directed exciton transport through long chain-like systems with an intrinsic on-site energy gradient~\cite{SD-JCP-biased-chains}. Bright states~\cite{OQS:super-absorption-realisation}, on the other hand, may offer advantages for faster conversion of photonic to electronic energy~\cite{OQS:Will-GuideSlide, OQS:Rouse-quantum-battery}.

The goal of any transport process is to achieve net energy flow from a point `A' to some spatially separated point `B'. Perhaps the simplest approach to achieving this goal is to introduce a net `downhill' energy gradient from A to B; therefore allowing thermal relaxation processes to mediate transport (similar `noise-assisted' processes also underlie the related phenomena of Environmentally-Assisted Quantum Transport~\cite{QT:OriginalENAQT, QT:Original-ish-ENAQT, QT:UniversalOriginENAQT, QT:ion-chain-ENAQT} and Vibration-Assisted Quantum Transport~\cite{QT:LovettVAET, OQS:OlayaCastroVAET, QT:Dubi-VAET-local-vs-global-phonons, QT:PhycobilsomeVAET, QT:TrappedIonVAET-PRX}). There are a variety of ways in which to introduce an energy gradient in nanoscale transport systems; e.g.~chemical substitution in molecular systems~\cite{QT:molecular-energy-grad}, local strain engineering in atomically thin 1D and 2D materials~\cite{QT:exciton-energy-grad1,QT:exciton-energy-grad2,QT:exciton-energy-grad3}, intrinsic electric fields within polar transition-metal-oxide heterostructures~\cite{QT:Electric-field-TMOs_1,QT:Electric-field-TMOs_2} and external electric fields applied to coupled quantum wells~\cite{QT:CQW-Energy-Gradient_1,QT:CQW-Energy-Gradient_2,QT:energy-grad-indirect-QD-excitons}. For simple proof of concept experiments, fine-grained control over energetic gradients, as well as other system parameters, can also be achieved with superconducting circuits~\cite{QT:super-cond-light-harv}. Furthermore, certain natural photosynthetic systems also feature an intrinsic energy gradient~\cite{QT:MomentInertia, QT:FMO-energy-grad, Intro:PhycobilisomeEnergyGrad-1, Intro:PhycobilisomeEnergyGrad-2}, and it has been suggested that this may be a key factor in determining transport efficiency~\cite{QT:Kassal-photosynth-energy-funnelling}.

In this work, we demonstrate that simultaneously efficient and robust long-range transport down an energetic gradient can be achieved for certain simple geometric arrangements of sites within a repeated chain of planar unit cells. First, in Sec.~\ref{sec:model}, we construct an open quantum systems model of excitonic energy transport which accounts for the effects of both electromagnetic (EM) and vibrational system-environment interactions -- both of which would likely be unavoidable in molecular or solid-state transport networks. From there, Sec.~\ref{sec:results} presents evidence to support the main results of this work; namely that, by appropriate choice of unit cell geometry, we can achieve transport efficiency which is effectively independent of transport distance while also being robust to relatively large energetic disorder. In Sec.~\ref{sec:eigenbasis}, we perform a detailed analysis of the physical mechanisms underlying this remarkable transport behaviour and show that it stems from a energetic separation of the bright and dark states within the system, allowing transport to proceed primarily through a low-energy `dark chain'. Finally, in Sec.~\ref{sec:dipoles}, we augment our model to include the relevant physics of intrinsic site dipoles and show that the appropriate choice of dipole orientations can lead to additional protection against both radiative and non-radiative losses.

\section{Transport Model}
\label{sec:model}

\begin{figure}
    \centering
    \includegraphics[scale=0.32]{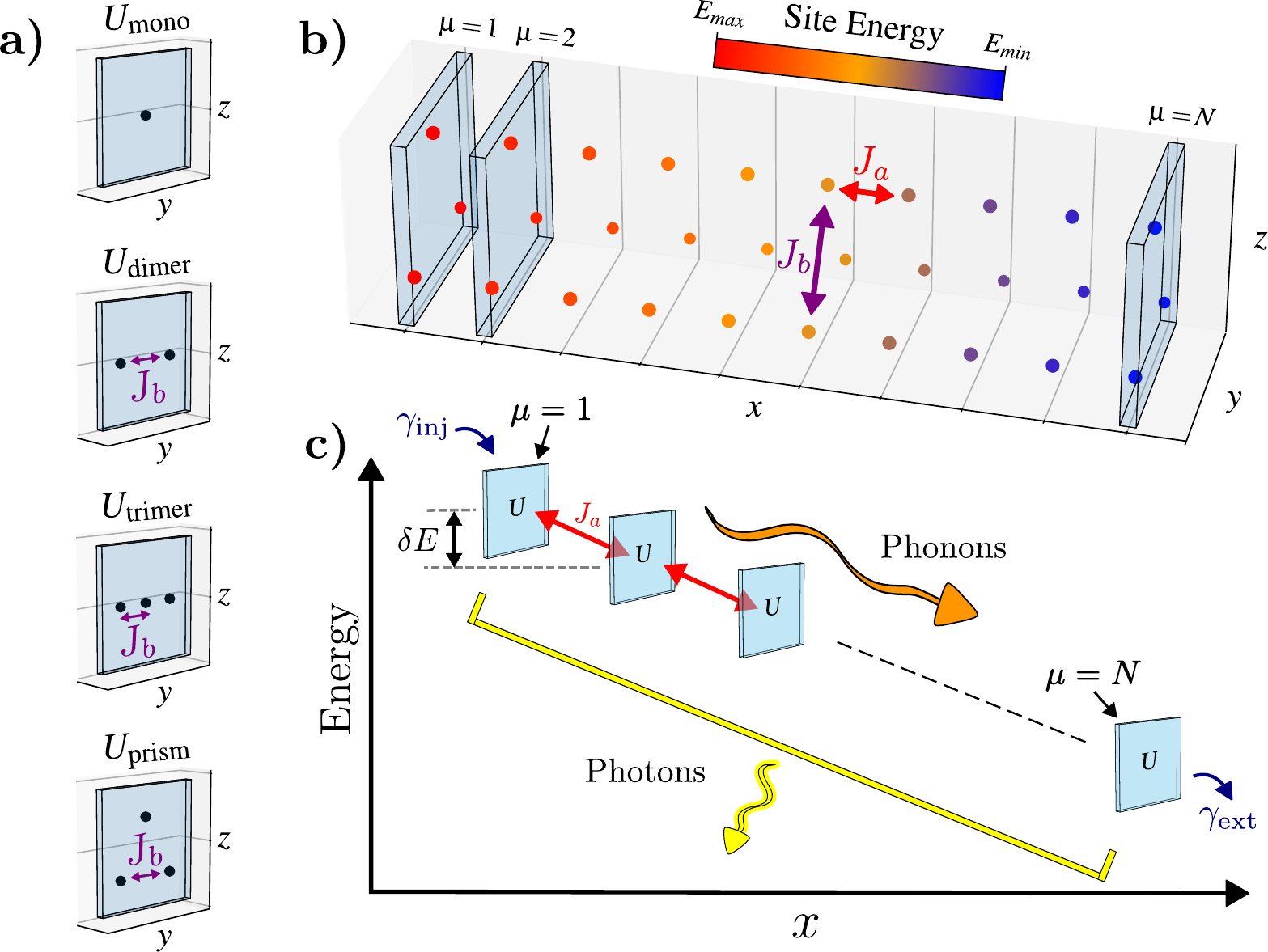}
    \caption{
        Depiction of the transport model used throughout this work with: \textbf{a)} the various unit-cell geometries considered (these panels are used as a key in all later plots); \textbf{b)} example 3D network structure ($U_\text{prism}$) with planar unit cells orthogonal to both the transport direction and the intrinsic energy gradient; \textbf{c)} simplified systems diagram illustrating the relevant physical processes and interactions included in our model. All sites within a single unit cell have identical site energy, with neighbouring unit cells detuned by energy $\delta E$. Throughout this work we use Greek indices ($\mu, \nu$) to denote different unit cells, and Latin indices ($i, j$) to distinguish sites within each unit cell.
    }
    \label{fig:system}
\end{figure}

To begin, we construct a minimal model of excitonic energy transport by following the established procedure of specifying a simple tight-binding Hamiltonian for the transport system, while modelling the effects of the ambient EM and vibrational environments using weak-coupling open quantum systems approaches~\cite{OQS:BreuerPetruccione, QT:LHC-Exciton-Review-Jang-2018, QT:TransportModels}. We restrict our model to the single-excitation subspace, which is well justified for realistic photosynthetic systems (since excitation events are rare) and for larger systems with an intrinsic energy gradient (since excitonic states remain relatively localized so exciton-exciton interactions can be neglected). Importantly, this approximation leads to a computationally efficient model which allows for the investigation of relatively large transport networks. On the other hand, this modelling approach rules out the study of charge transport since it neglects multi-exciton states and assumes that the electron and hole constituents of the exciton are completely bound. As such, we restrict our discussion to excitonic energy transport systems throughout this work.

The Hamiltonian for the excitonic part of the system, depicted in Fig.~\ref{fig:system}, takes the form
\begin{equation}
\label{eq:H-sys}
\begin{split}
    \hat{H}_{s} = &\sum_{\mu=1}^N \sum_{i=1}^n \big[ (N-\mu) \delta E + E_0 \big] \outerproduct{\mu, i}{\mu, i} + E_g \outerproduct{g}{g} \\ 
    &+ \sum_{\mu \neq \nu, i, j} \frac{J_a}{|\vec{r}_{\mu, i} - \vec{r}_{\nu, j}|^3} \outerproduct{\mu, i}{\nu, j} \\
    &+ \sum_{\mu, i \neq j} \frac{J_b}{|\vec{r}_{\mu, i} - \vec{r}_{\mu, j}|^3} \outerproduct{\mu, i}{\mu, j} ~,
\end{split}
\end{equation}
where $N$ is the number of unit cells, $n$ is the number of sites per unit cell, $\ket{\mu, i}$ denotes a localized site-basis state on the $i^{th}$ site of unit cell $\mu$, $\delta E$ is the on-site energy detuning between neighbouring unit cells, $E_0$ and $E_g$ set the energy difference between the ground state $\ket{g}$ and the excited state manifold and $\vec{r}_{\mu, i}$ is the real space position of site $\ket{\mu, i}$. We emphasize here that the coherent coupling terms in Eq.~\eqref{eq:H-sys} are `all to all' couplings where each site is coupled to all other sites, with a coupling strength proportional to the inverse separation cubed. As such, the values of $J_a$ and $J_b$ determine the \textit{relative} strength of inter- vs intra- unit cell coupling between respective nearest neighbours. Furthermore, by modelling the hopping terms with a $1/|r|^3$ distance dependence, we implicitly focus on resonant F\"orster transport, typically associated with singlet excitons. However, the exponential distance dependence which appears in Dexter-coupled triplet exciton transfer~\cite{Intro:triplet-exp-hopping} should only lead to minor variations in Hamiltonian eigenstate structure. As such, we expect that the core results presented in this work will hold for different mathematical forms of the exciton hopping terms.

Throughout the rest of this work, we fix $J_a = 1$ which, as explained in detail in Appendix~\ref{apdx:dimensionless-model}, allows us to express the model parameters in a dimensionless form, where all quantities are relative to the maximal inter-unit-cell coupling strength $J_a$. This allows us to focus on relative parameter scales without constraining our model to the parameter values of relevance to a specific physical system.

In order to model the interaction between our system and the external environment within which it is embedded, we use a weak-coupling Pauli Master Equation (PME) to describe the dynamical evolution of the system eigenstate populations ($P_n$). The PME takes the form 
\begin{equation}
    \label{eq:PME}
    \partial_t P_n = \sum_m \left[ W_{nm} P_m(t) - W_{mn} P_n(t) \right] ~,
\end{equation}
where the matrix element $W_{nm}$ contains the total transition rate from eigenstate $\ket{\phi_m}$ into eigenstate $\ket{\phi_n}$ and is given by
\begin{equation}
    \label{eq:W-matrix}
    W_{nm} = \sum_\alpha S_\alpha(\omega_{mn}) \matrixelement{\phi_m}{\hat{A}_\alpha}{\phi_n} \!\!\! \matrixelement{\phi_n}{\hat{A}_\alpha}{\phi_m} ~,
\end{equation}
which in turn depends on the spectral density $S_\alpha$ of each environment $\alpha$, evaluated at the eigenenergy difference $\omega_{mn} = \varepsilon_m - \varepsilon_n$, as well as the system part of the system-environment interaction Hamiltonian $\hat{H}_I = \sum_\alpha \hat{A}_\alpha \otimes \hat{B}_\alpha$. Although the PME is a dynamical equation, we are only concerned with steady state solutions to Eq.~\eqref{eq:PME} (obtained from the vector form $\partial_t \vec{P} = \chi \vec{P}$ via the null-space of the matrix $\chi$). The strong agreement between our PME model and a more accurate Bloch-Redfield master equation (see Fig.~\ref{fig:multi-chain} and Appendix~\ref{apdx:BRME}) further justifies our use of the simple PME description.

We model the vibrational environment surrounding our system as set of independent site-local phonon baths, with system operator $\hat{A}_\text{(vib, $\mu$, i)} = \outerproduct{\mu, i}{\mu, i}$.
The phonon bath density of states is modelled using a Drude-Lorentz spectral density; as is common in excitonic transport systems, where the spectrum falls to zero in both the high and low frequency limits, while exhibiting a single peak in between. The Drude-Lorentz spectrum is given by
\begin{equation}
    \label{eq:Sw-DL}
    S_{DL}(\omega) = \frac{\pi |\omega| \Gamma \gamma_{ph}}{\Gamma^2 + (|\omega| - \omega_0)^2} \left[n_{BE}(|\omega|, T_{ph}) + \Theta(\omega)\right] ~,
\end{equation}
where $\gamma_{ph}$ controls the overall system-vibration coupling, $\Gamma$ determines the width of the phonon spectrum,  $\omega_0$ determines the peak location and $n_{BE}$ is the Bose-Einstein distribution at temperature $T_{ph}$. The Heaviside step function $\Theta(\omega)$ accounts for thermodynamic detailed balance and ensures relaxation to the correct thermal (Gibbs) state in the absence of other system-environment interactions. In Appendix~\ref{apdx:phonon-param-disorder}, we show that the key results presented in this work are valid across a range of Drude-Lorentz spectrum parameters $\gamma_{ph}$, $\Gamma$ and $\omega_0$. It is worth noting that, while the Drude-Lorentz spectrum used here is derived from an over-damped Brownian oscillator model and is broadly applicable~\cite{QT:FMO-energy-grad}, specific physical systems in which the (principal) phonon modes of the environment are known may lend themselves to alternate spectral densities which capture the vibrational properties of the material in question more accurately.

For all other system-environment interactions we use uni-directional Lindblad-like operators ($\hat{\mathcal{A}}$) which are constructed by combining the appropriate (Hermitian) $\hat{A}$ operator with a spectral density whose only $\omega$ dependence is in the Heaviside step function $\Theta(\omega)$ (full details in Appendix~\ref{apdx:lindblad-processes}).

With this in mind, we model radiative recombination processes, which occur due to (collective) interactions with the ambient EM field, using an operator of the form
\begin{equation}
    \label{eq:A-radiative}
    \hat{\mathcal{A}}_\text{rad} = \sum_{\mu, i} \ \outerproduct{g}{\mu, i} ~,
\end{equation}
with an associated rate $\gamma_\text{rad}$ which determines the overall coupling between system and electromagnetic environment. Physically, this form of interaction assumes that the localization length of each system eigenstate is much less than the wavelength of the ambient (optical) photons, such that all sites within the exciton state interact in-phase with the EM field. This assumption is justified due to the intrinsic energy gradient in our system which causes eigenstates to remain relatively localized (see Fig.~\ref{fig:single-chain}). The use of a uni-directional operator, i.e. one which does not account for photon re-absorption (negligible at typical nanoscopic energy scales and ambient temperatures), is justified here since we are primarily concerned with the effects of detrimental radiative \textit{loss} during transport.

In addition to the electromagnetic and vibrational environments discussed above, we also include a set of phenomenological injection processes which generate excitations on each site $i$ within the unit cell $\mu = 1$, to simulate the initial excitation process which must occur before transport; as well as a corresponding set of extraction processes (from each site $i$ within $\mu = N$) to mimic the eventual extraction of energy for conversion into useful work (e.g.~to initiate charge separation in photosynthetic systems). These are included as
\begin{align}
    \hat{\mathcal{A}}_\text{inj, i} &= \outerproduct{1, i}{g} ~, \label{eq:A-inj} \\ 
    \hat{\mathcal{A}}_\text{ext, i} &= \outerproduct{g}{N, i} ~, \label{eq:A-ext}
\end{align}
with associated injection and extraction rates $\gamma_\text{inj}$ and $\gamma_\text{ext}$ respectively.

While most realistic systems will suffer from both radiative and non-radiative losses, our primary focus with this work is mitigating radiative loss; therefore we do not explicitly model non-radiative loss at this stage. However, in Sec.~\ref{sec:dipoles} we modify our model to consider the effects of the intrinsic dipole moments which will be present in certain physical systems, and discuss how this can also lead to a reduction in non-radiative losses.

Finally, in order to quantify transport efficiency we use the steady state exciton current flowing through the system, defined as
\begin{equation}
    \label{eq:ss-current}
    I_{ss} = \sum_i \gamma_\text{ext} \bra{N, i} \hat{\rho}_{ss} \ket{N, i},
\end{equation}
where $\hat{\rho}_{ss}$ is the steady state density matrix of the system. The sum over $i$ signifies that we include the total current from all sites in the final unit cell ($\mu = N$) of the chain.

\section{Results: efficient \& robust transport}
\label{sec:results}

\subsection{Single Chain Transport Incurs Losses}
\label{sec:single-chain}

\begin{figure}
\begin{flushleft}
    \includegraphics[scale=0.44]{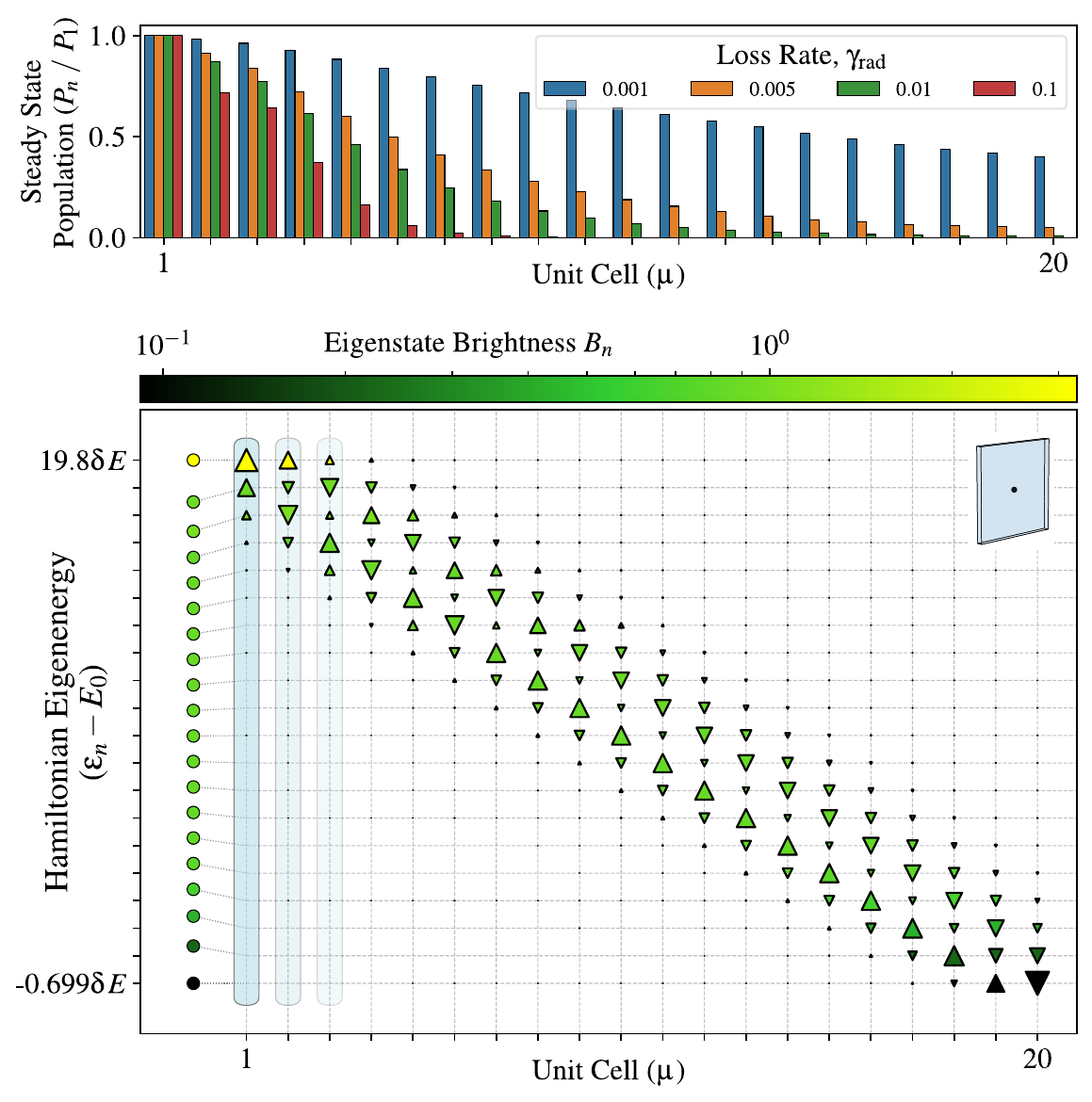}
    \caption{
        Summary of the transport properties of a single linear chain ($U_\text{mono}$) with intrinsic energy gradient. \textit{Top} -- Individual site populations (relative to $P_1$) decrease with increasing $\gamma_\text{rad}$. \textit{Bottom} -- Eigenbasis structure of the single-chain system Hamiltonian. Each row of triangular markers represents a single eigenstate, with marker size indicating magnitude, and upwards (downwards) triangle orientation indicating positive (negative) phase, of the eigenstate's site-basis components. Circular markers on left indicate corresponding eigenenergies and colour scale denotes optical brightness [Eq.~\eqref{eq:brightness}].
    }
    \label{fig:single-chain}
\end{flushleft}
\end{figure}

We begin our investigation by considering the simplest possible transport system encompassed by our generic model -- the linear chain with a single site per unit cell (\cell{mono} in Fig.~\ref{fig:system}). To allow for a fair comparison between this system and the more complex geometries discussed later, we choose the phonon bath parameters such that the temperature independent part of Eq.~\eqref{eq:Sw-DL} is optimal for transport with \cell{mono}. This means that we set $\omega_0^2 = \delta E ^2 - \Gamma^2$ (derived by neglecting the term in square brackets, solving $\frac{d S_{DL}}{d\omega} = 0$ for $\omega_0$ then setting $\omega = \delta E$) so that fast phonon-mediated transitions occur between eigenstates separated by energy difference $\delta E$. This choice will favour the single site unit cell configuration, and thus provide a fair benchmark for comparing the multi-site unit cell cases below. We maintain this same $\omega_0$ value for all other unit cells throughout this work

As shown in Fig.~\ref{fig:single-chain}, despite these optimal phonon parameters, the linear chain network geometry is highly sensitive to radiative loss rate $\gamma_\text{rad}$, with a gradual reduction in (relative) population on the sites near the bottom of the chain as $\gamma_\text{rad}$ increases. This failure to transport population down the chain can be explained by considering the bottom panel of Fig.~\ref{fig:single-chain}, which illustrates the eigenstate structure of the system Hamiltonian. Firstly, this panel demonstrates that the eigenstates remain relatively localized over only a few sites and that the corresponding eigenenergies are approximately evenly spaced (excluding boundary effects). Secondly, by examining the \textit{optical brightness}, defined as 
\begin{equation}
    \label{eq:brightness}
    B_{n} = |\matrixelement{g}{\hat{\mathcal{A}}_\text{rad}}{\phi_n}|^2 ~,
\end{equation}
for each excitonic eigenstate of the system, we can begin to understand the poor transport performance of this single chain geometry. Specifically, the relatively uniform brightness of eigenstates within the bulk of the system results in significant radiative loss during transport, and therefore very little population reaching the bottom of the chain and contributing to the steady state current. In other words, excitons are transported via a sequence of states from which they can easily recombine and radiate away their energy.

We also find that this single chain system exhibits particularly poor scaling of steady state current with chain length (see Fig.~\ref{fig:multi-chain}), with a steep exponential decrease in current for longer chains. In the next section, we show that this scaling can be dramatically improved by considering systems with multiple sites per unit cell and tuning the value of $J_b$ appropriately.

\subsection{Multi-Site Unit Cells Reduce Losses}
\label{sec:multi-chain}

\begin{figure*}
    \includegraphics[scale=0.59]{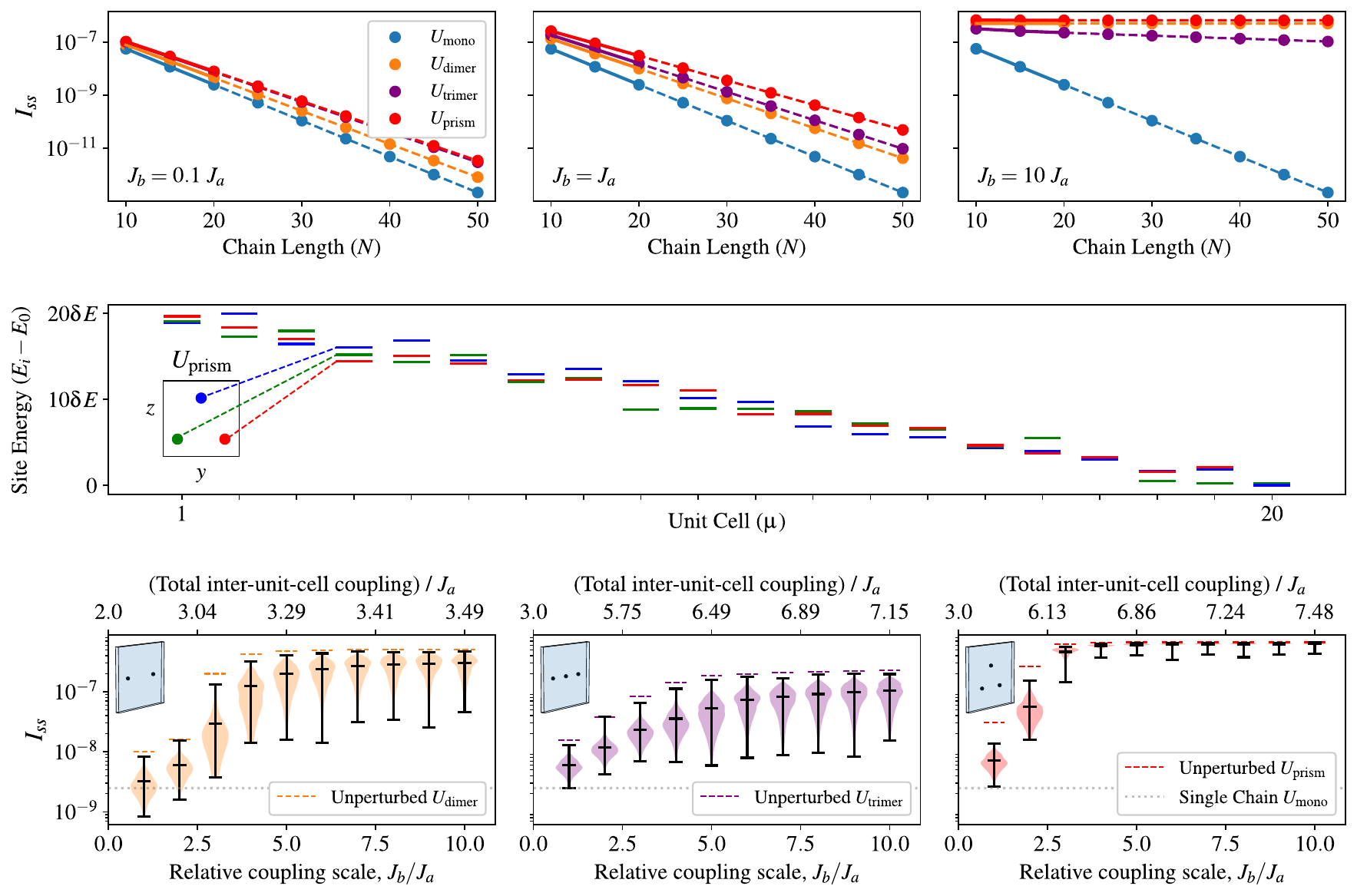}
    \caption{
        Comparison of transport properties for different unit cell geometries and $J_b$ values. \textit{Top} -- Scaling of transport efficiency with chain length. Marginal performance improvements are seen for the multi-site unit cells in the $J_b \leq J_a$ regimes; significant performance improvements are seen the $J_b > J_a$ regime, with $I_{ss}$ becoming effectively independent of chain length for the $U_\text{dimer}$ \& $U_\text{prism}$ cases. Solid lines up to chain length 20 in all three panels show results obtained with a non-secular Bloch-Redfield master equation (see Appendix~\ref{apdx:BRME}). \textit{Middle} -- On-site energies for a single realization of the disordered Hamiltonian (Eq.~\eqref{eq:H-dis}) with $\sigma = 0.9 \delta E$. \textit{Bottom} -- Robustness of transport efficiency to on-site energy perturbations for chains of 20 unit cells with $\sigma = 0.9 \delta E$. Both $U_\text{dimer}$ \& $U_\text{trimer}$ cases are significantly better than the single chain but exhibit large fluctuations in current when subject to on-site energy perturbations. The $U_\text{prism}$ geometry provides both a larger current in the optimal cases, and far better robustness at large $J_b$ values. The secondary x-axis scale on the bottom three panels shows the total inter-unit-cell coupling (in units of $J_a$) between neighbouring unit cells.
    }
    \label{fig:multi-chain}
\end{figure*}

In previous work, we found that the performance of a single chain transport system could be greatly (and counter-intuitively) improved by introducing energetic `barriers' at regular intervals along the chain~\cite{SD-JCP-biased-chains}. This resulted in the formation of evenly spaced (in energy) dark states and allowed transport to proceed via phonon-mediated hopping between these dark states; thus improving transport performance significantly. In certain experimental implementations, introducing the aforementioned energetic `barriers' by manipulating individual site energies in a chain may be challenging. In this section, we show that extremely large enhancements in transport efficiency may be achieved by introducing alternative unit cell geometries, while keeping the simple linear energy gradient intact. 

For simplicity, we focus on the three multi-site unit cell geometries, \cell{dimer}, \cell{trimer} and \cell{prism}, illustrated in Fig.~\ref{fig:system} -- all of which can be equivalently thought of as multiple `single chain' systems arranged in parallel, with nearest neighbour inter-\textit{chain} coupling $J_b$. It is worth noting that, due to the pair-wise coupling in Eq.~\eqref{eq:H-sys}, the \textit{total} coherent coupling between multi-site unit cell geometries will be larger than the \cell{mono} case; however, for ease of comparison we focus on the ratio $J_b/J_a$ rather than the total ratio of intra- to inter-unit-cell couplings throughout most of this work (aside from Fig.~\ref{fig:multi-chain} where we consider both). Furthermore, to allow for fair comparison between the single and multi-site unit cells, we rescale the system injection rate $\gamma_\text{inj} \mapsto \gamma_\text{inj} / n$ (where $n$ is the number of sites per unit cell) in order to keep the total excitation rate constant in all cases. This ensures that all unit cell geometries generate the same steady state current in the limit $J_b \rightarrow 0$.

In the top row of Fig.~\ref{fig:multi-chain} we compare the transport performance of the single chain system with that of each multi-site unit cell configuration at three different values of $J_b$. We find that in all three regimes ($J_b < J_a$, $J_b = J_a$ and $J_b > J_a$) the multi-site unit cells perform better than the single chain case, with larger enhancements for longer range transport. This is particularly surprising since the overall oscillator strength of the system is proportional to the number of system sites.

Furthermore, in the $J_b > J_a$ regime (top right panel Fig.~\ref{fig:multi-chain}) we find a sharp change in the performance of the multi-site unit cell cases. Most strikingly, the $U_\text{dimer}$ and $U_\text{prism}$ cases are almost perfectly flat; suggesting that, in these two cases at least, steady state exciton current is effectively \textit{independent} of chain length. By performing an exponential least-squares fit to the function $I_{ss} = \alpha e^{-\beta \cdot N}$ (where $N$ is the chain length) for the data in Fig.~\ref{fig:multi-chain}, we find that the drop off in current as a function of chain length for the $U_\text{prism}$ case changes from $\beta \approx 0.26$ for $J_b = 0.1\,J_a$ to as little as $\beta \approx 1.2 \times 10^{-4}$ when $J_b = 10\,J_a$. In contrast, we find $\beta \approx 0.31$ for the $U_\text{mono}$ case. The origin of this effect will be analysed and explained in detail in Sec.~\ref{sec:discussion}.

\subsection{Transport Robustness Despite Disorder}
\label{sec:robust-to-Es}
The eigenstates of a degenerate two-site system are perfectly delocalized (bonding/anti-bonding) states for all non-zero values of inter-site coupling, whereas this perfect delocalization is broken for any infinitesimal degeneracy-breaking energy perturbation. Therefore, it is reasonable to question whether the remarkable improvements in transport efficiency revealed in the previous section are reliant on maintaining the perfect degeneracy of all sites within each unit cell.

To investigate this question, we add random degeneracy-breaking perturbations to each on-site energy in our system Hamiltonian to obtain a modified Hamiltonian $\tilde{H}_s$ given by
\begin{equation}
    \label{eq:H-dis}
    \tilde{H}_s = H_s + \sum_{\mu, i}\Delta_{\mu, i}(\sigma) \outerproduct{\mu, i}{\mu, i}
\end{equation}
where each energy perturbation $\Delta_{\mu, i}$ is sampled from a Gaussian distribution with zero-mean and standard deviation $\sigma$. The middle panel of Fig.~\ref{fig:multi-chain} contains a single illustrative example of the perturbed system energies for the modified Hamiltonian $\tilde{H}_s$ with $\sigma = 0.9 \delta E$. Disorder of this magnitude is sufficient to break the `downhill' ordering of the site energies and should intuitively inhibit phonon-mediated transport.

The bottom row of Fig.~\ref{fig:multi-chain} shows the distribution of steady state currents for each of the multi-site unit cell systems over a range of $J_b$ values with $10^3$ realizations of the disordered Hamiltonian $\tilde{H}_s$ at each $J_b$. We observe that, in agreement with Sec.~\ref{sec:multi-chain}, transport performance is orders of magnitude better in the $J_b \gg J_a$ regime on average, despite the addition of energetic disorder. Furthermore, we find that the transport performance of the triangular prism geometry $U_\text{prism}$ is remarkably robust, even for the relatively large $\sigma = 0.9 \delta E$ disorder used here.

In this Section, we have shown that robust \textit{and} efficient long-range exciton transport can be achieved in systems containing multiple sites per unit cell by tuning the system parameters such that the intra-unit-cell coupling ($J_b$) is much larger than the inter-unit-cell coupling ($J_a$). In the next section, we reveal the surprisingly simple physics underlying this transport behaviour by analysing the relevant eigenbasis properties of the system Hamiltonian.

\section{Discussion: Underlying Physics}
\label{sec:discussion}

\subsection{Eigenbasis Bipartition \& Dark Chain Transport}
\label{sec:eigenbasis}

\begin{figure*}
    \includegraphics[scale=0.5]{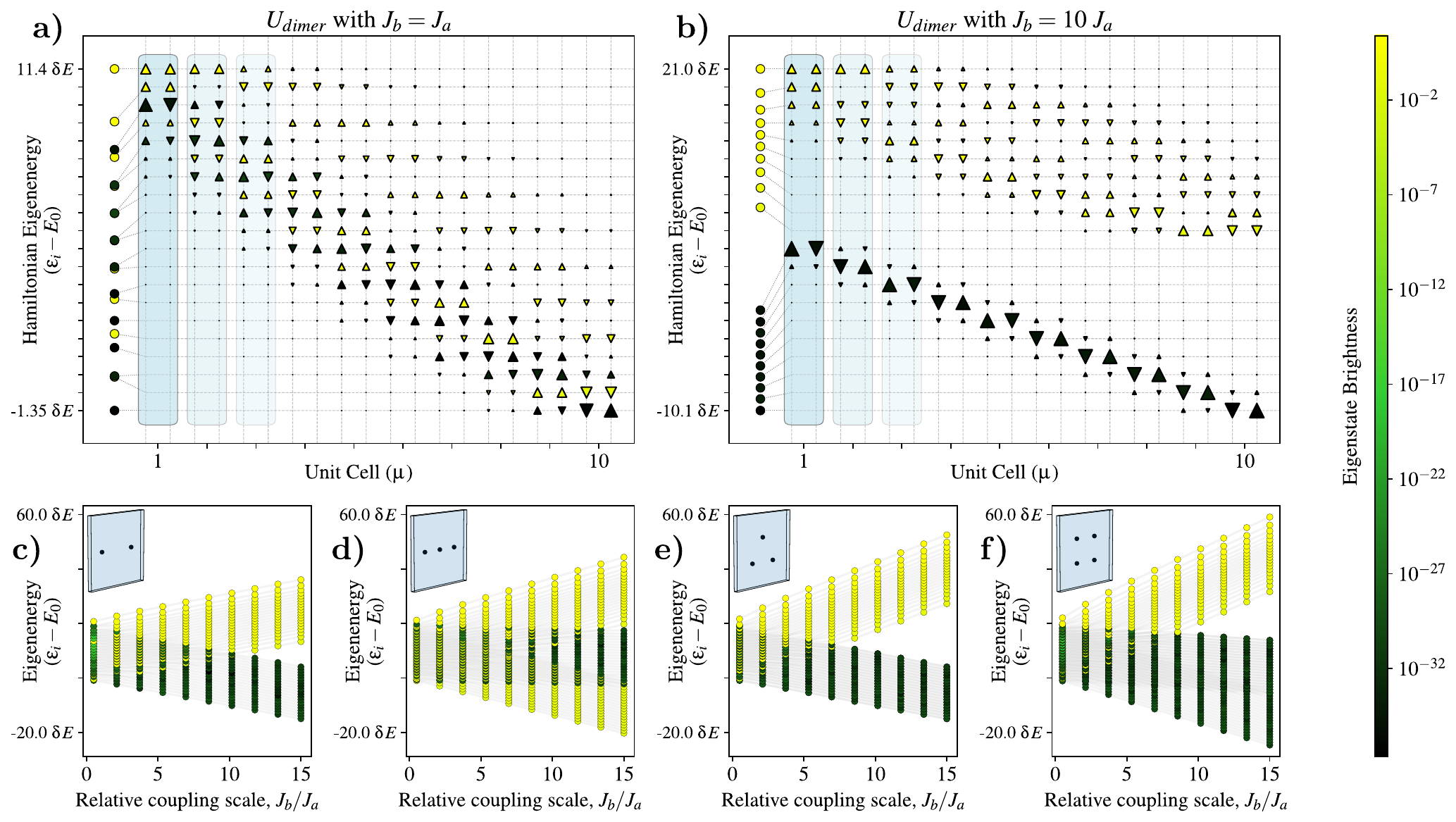}
    \caption{
        Eigenbasis properties of systems with different unit-cell types and $J_b$ values. Panel \textbf{a)} spatial eigenstate structure for $U_\text{dimer}$ with alternating bright and dark states along the length of the chain (see caption of Fig.~\ref{fig:single-chain} for an explanation of the plot format); \textbf{b)}  spatial eigenstate structure for $J_b \gg J_a$ where the eigenstates form two energetically separated chains, with phonon-mediated transport through the lower energy `dark chain' providing protection from radiative losses; \textbf{c)} to \textbf{f)} eigenenergy spectra vs $J_b$ for chains of 20 unit cells, showing the increasing energetic separation between bright and dark states, with dark states at lower energy in all except the $U_\text{trimer}$ case. \textit{Note} -- the largest brightness value in each panel is subtly different, but these variations are indistinguishable on the broad logarithmic colour scale.
    }
    \label{fig:eigenbasis}
\end{figure*}

As discussed in Sec.~\ref{sec:single-chain}, the single-chain transport system exhibits a simple eigenstate structure, with relatively uniform optical brightness throughout the chain leading to poor transport performance. Fig.~\ref{fig:eigenbasis}a shows the contrasting eigenstate structure of the $U_\text{dimer}$ case with $J_b = J_a$. Here, we see a very different pattern of alternating bright and dark states along the chain with a far greater range of brightness values spanned in the $U_\text{dimer}$ case compared with the $U_\text{mono}$ case (as seen by comparing the colour scales in Figs.~\ref{fig:single-chain}~\&~\ref{fig:eigenbasis}) leading to darker eigenstates which provide significantly better protection from radiative recombination in the $U_\text{dimer}$ case. Furthermore, as derived in Appendix~E of~\cite{SD-JCP-biased-chains}, the phonon-mediated transition rate between between two eigenstates depends on both the mutual site-basis overlap and the pairwise eigenenergy difference between states. Therefore, the presence of delocalized `tails' of site-basis support stretching down the chain for each of the bright states in Fig.~\ref{fig:eigenbasis}a, compared with the relatively localised dark states, will reduce the rates of transitions between dark and bright states. This observation, combined with the approximately uniform separation of dark state eigenenergies, leads to the dominance of phonon-mediated dark state to dark state transitions, thereby minimizing the populations of bright eigenstates and reducing the likelihood of detrimental radiative losses (see Appendix~\ref{apdx:phonon-rates} for more details).

Fig.~\ref{fig:eigenbasis}b contains the same eigenstate structure plot format for the $U_\text{dimer}$ case in the $J_b \gg J_a$ regime, which allows us to better understand the remarkable transport behaviour observed in Fig.~\ref{fig:multi-chain}. In this regime, the bright and dark eigenstates become energetically separated, with the dark states forming a lower energy `dark chain' which spans the complete length of the system and which is almost completely immune to radiative loss. The presence of a realistic finite-temperature vibrational environment in our model leads to thermal relaxation processes which preferentially funnel excitations towards the lower energy eigenstates. Therefore, when these low energy eigenstates form the aforementioned dark chain, as also happens in the $U_\text{prism}$ case (see Appendix~\ref{apdx:prism-eigenstructure}), the long distance transport with minimal radiative loss seen in Fig.~\ref{fig:multi-chain} can be understood as arising from this relatively simple dark state protection scheme.

Although both $U_\text{dimer}$ and $U_\text{prism}$ geometries demonstrate the required eigenstate structure for dark chain transport at large $J_b$, the current flowing through the triangular prism system was shown to be significantly more robust to disorder in the previous section. This can be explained by noting that the dimer system has an equal number of bright and dark states, whereas the triangular prism has twice as many dark states as bright states. This observation, combined with the fact that the \cell{prism} case exhibits less mixing between bright and dark eigenstates when subject to on-site energy perturbations (see Appendix.~\ref{apdx:prism-robustness}), explains the observed differences in transport robustness.

The general increase in transport efficiency at larger $J_b$ values, as observed in Fig.~\ref{fig:multi-chain}, can be explained by considering Fig.~\ref{fig:eigenbasis}c--f. Here, we see the different energetic trajectories of bright vs dark states as a function of $J_b$, with the gradual formation of the aforementioned low energy `dark chain' in the $U_\text{dimer}$ and $U_\text{prism}$ cases. This same analysis also explains the slightly poorer performance of the $U_\text{trimer}$ case observed in Fig.~\ref{fig:multi-chain} since, as shown in Fig.~\ref{fig:eigenbasis}d, the eigenstates instead separate into three distinct subsets, with the lowest energy subset being brighter than the middle subset. Therefore, when thermal relaxation processes funnel excitations into the low energy eigenstates, the system becomes more susceptible to radiative loss, leading to poorer transport performance for long chains.

Additionally, comparing Fig.~\ref{fig:eigenbasis}e and \ref{fig:eigenbasis}f reveals that increasing the number of sites per unit cell in a prism arrangement (e.g. the cuboid arrangement of Fig.~\ref{fig:eigenbasis}f) is unlikely to lead to further improvements compared with $U_\text{prism}$ since the inclusion of extra sites decreases the energetic separation between bright and dark eigenstate band, and will therefore increase bright state populations. In any practical system it may also be desirable to minimize the number of system sites the transport network comprises, due to physical resource constraints, or to simplify the experimental implementation. This will be particularly important in systems where \textit{non}-radiative decay processes are prevalent, since minimizing the time each excitation spends in the network becomes crucial.

Finally, it is worth noting that in order to utilize the `dark chain' transport scheme elucidated in this section, the system must first populate the dark states near the top of the chain. By definition, this cannot be achieved directly through photon absorption. Due to our chosen form of injection operator (Eq.~\eqref{eq:A-inj}) -- which is equivalent to excitation via near-field F\"orster coupling -- these dark states can become populated through direct injection as well as through phonon-mediated eigenstate transitions from the higher energy bright states. In certain physical systems where initial excitation is achieved via the absorption of an optical photon, a more accurate model of the injection process should instead populate some mixture of high energy bright states. The dark chain transport scheme would then rely more heavily on phonon processes funnelling population into the highest energy dark states; however, as shown in Appendix~\ref{apdx:eigen-inj}, this leads to only minor variations in transport efficiency. Furthermore, this limitation can often be overcome in practice via the use of sensitizer quantum dots or molecules as part of the initial excitation mechanism~\cite{QT:SolarCell-QD-sensitizer}.

In the next section, we add site dipoles and non-radiative recombination to our model -- in order to more accurately describe the physics of certain relevant experimental platforms, such as molecular networks -- and show that these changes can lead to further interesting transport phenomena.

\subsection{Dipole Effects Mitigate Non-radiative Losses}
\label{sec:dipoles}

\begin{figure*}
    \includegraphics[scale=0.68]{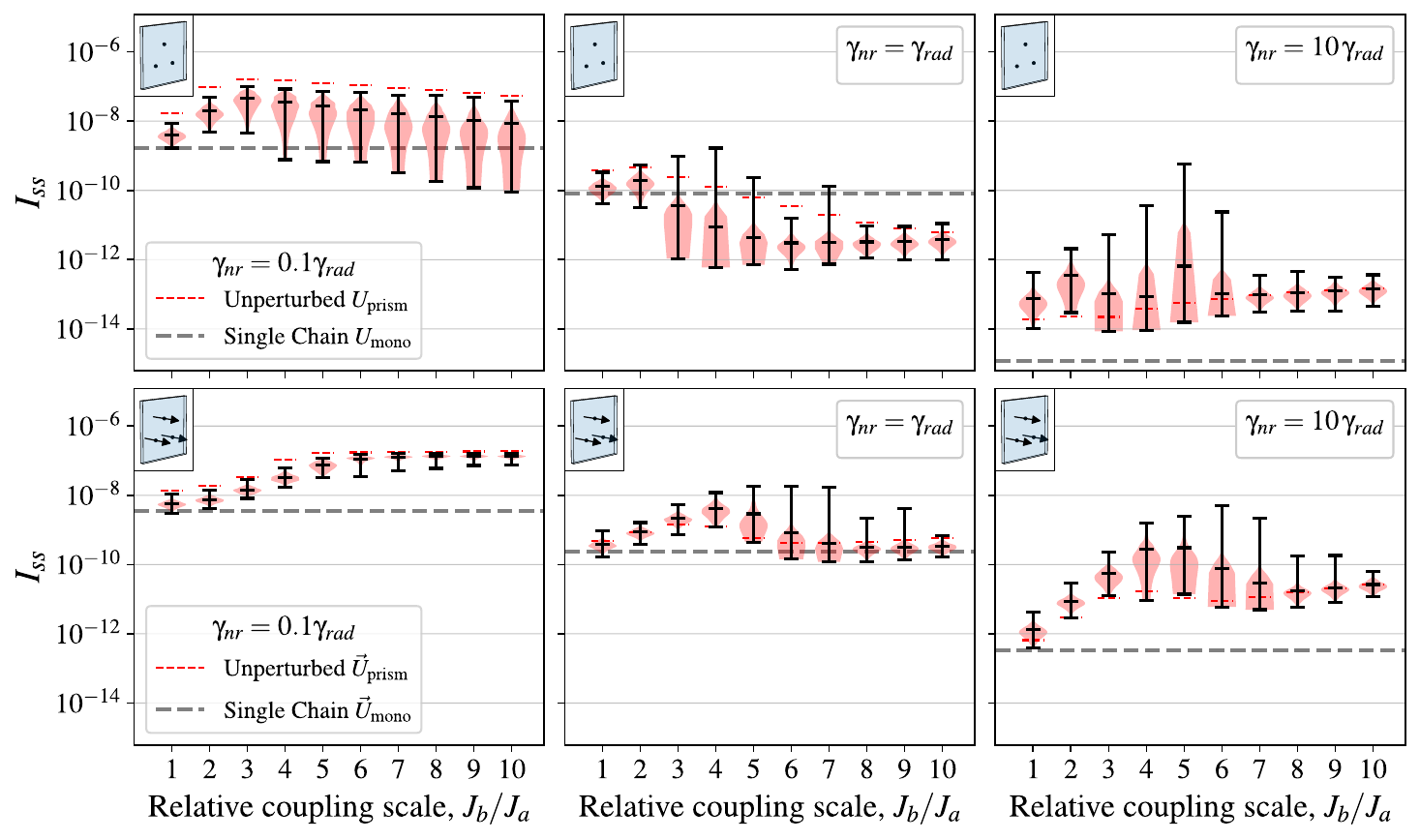}
    \caption{
        Transport efficiency and robustness of $U_\text{prism}$ to on-site energy perturbations as a function of $J_b$ at three different $\gamma_\text{nr}$ values. (\textit{Top left}) -- transport performance for the perturbed cases can often be worse than the (unperturbed) single chain even with a relatively small non-radiative loss rate; (\textit{top middle}) -- upon increasing $\gamma_\text{nr}$ the majority of perturbed $U_\text{prism}$ cases perform worse than the (unperturbed) single chain;  (\textit{top right}) -- increasing $\gamma_\text{nr}$ leads to even poorer transport for both prism and single chain; (\textit{bottom row}) -- by including the effects of intrinsic site dipoles and aligning these dipoles along the direction of transport (see insets), the triangular prism geometry \dcell{prism} can recover its efficiency and robustness advantages over the single chain (see text for details). In all panels the radiative decay rate is fixed at $\gamma_\text{rad} = 0.01$ while $\gamma_\text{nr}$ is varied.
    }
    \label{fig:dipoles}
\end{figure*}

As mentioned in Sec.~\ref{sec:intro}, many physical systems to which our model is relevant will suffer from additional non-radiative recombination processes during transport which are, in principle, avoidable but are, in practice, often dominant over the (intrinsically unavoidable) radiative loss mechanisms examined thus far. Since non-radiative leakage rates may span many orders of magnitudes depending on the physical system in question, we explicitly study three different regimes here ($\gamma_\text{nr} \ll \gamma_\text{rad}$, $\gamma_\text{nr} \approx \gamma_\text{rad}$ \& $\gamma_\text{nr} \gg \gamma_\text{rad}$).

A minimal model for these extra loss channels may be introduced via the site-local phenomenological operators
\begin{equation}
    \mathcal{\hat{A}}_\text{(nr, $\mu$, i)} = \outerproduct{g}{\mu, i} ~,
\end{equation}
with associated rate $\gamma_{nr}$ denoting the non-radiative loss rate from the $i^{th}$ site of unit cell $\mu$. Since these processes are individual to each site (rather than collective effects), they indiscriminately penalize slow transport, regardless of the eigenstate brightness distribution within the system. 

Another relevant physical feature which we have thus far neglected in our model is the intrinsic dipole moment which each system site may possess (e.g.~due to the polar nature of individual molecules). To investigate the effects of these dipole moments, we can modify the all-to-all coherent inter-site coupling in our system Hamiltonian to instead take the form
\begin{equation}
    \label{eq:dipole-coupling}
    V_{\mu \nu i j} = J_{a/b} \bigg(\frac{\vec{d}_{\mu,i} \cdot \vec{d}_{\nu,j}}{|\vec{r}_{\mu,i} - \vec{r}_{\nu,j}|^3} - 3 \frac{ ( \vec{r}_{\mu,i} \cdot \vec{d}_{\nu,j} ) ( \vec{d}_{\mu,i} \cdot \vec{r}_{\nu,j} )}{|\vec{r}_{\mu,i} - \vec{r}_{\nu,j}|^5}\bigg) ~,
\end{equation}
where the pre-factor is $J_a$ for inter-unit-cell terms (i.e. $\mu \neq \nu$) and $J_b$ otherwise (i.e. when $\mu = \nu$).

We also modify our description of the radiative decay process to be
\begin{equation}
    \hat{\mathcal{A}}_\text{rad} = \sum_{\mu, i} \vec{d}_{\mu,i} \outerproduct{g}{\mu, i} ~,
\end{equation}
where $\vec{d}_{\mu,i}$ is the (normalized) dipole moment of site $\ket{\mu,i}$ and all other symbols have the same meaning as Sec.~\ref{sec:model}. These modifications lead to changes in both the spatial structure of the system eigenstates and, through Eq.~\eqref{eq:brightness}, the eigenstate brightness distribution -- both of which will affect the transport properties of the system.

Surprisingly, we find that the detrimental effects of non-radiative recombination can be at least partially mitigated through the appropriate choice of site dipole orientations $\vec{d}_{\mu, i}$ and unit cell geometry. Specifically, by aligning all dipoles parallel to the transport direction (i.e.~along the $x$-axis in Fig.~\ref{fig:system}), both the \dcell{mono} and \dcell{prism} unit cells outperform the equivalent non-dipole-dependent cases across all three $\gamma_\text{nr}$ regimes. In contrast, this same dipole arrangement for \dcell{dimer} \& \dcell{trimer} does not lead to similar improvements in transport performance (see Appendix~\ref{apdx:sheets-and-dipoles}). 

It is worth noting that the dipole-independent coupling used prior to this section leads to H-aggregate-like behaviour in the system; however, the choice of dipole orientation in this section leads to positive intra-unit-cell coupling values $V$ (i.e. H-aggregate-like behaviour within each unit cell) and negative coupling values between sites in different unit cells (J-aggregate-like behaviour across unit cells)~\cite{QT:HJ-aggregate-review}.

Before discussing the intricacies of the transport performance in various non-radiative decay rate regimes, it is important to emphasize that the non-radiative processes described here are completely unaffected by the optical brightness of each eigenstate. The only possible approach to minimizing non-radiative loss in our model is to transport the excitons through the system as quickly as possible, thereby minimizing the time window during which non-radiative recombination can occur. In practice, this requires that phonon-mediated eigenstate transitions happen as quickly and uniformly as possible, since even a single bottleneck (i.e. a slow transition between two eigenstates) can drastically increase losses and harm transport performance. With this in mind, we shall now discuss the three important parameter regimes for $\gamma_\text{nr}$.

In the first regime, when $\gamma_\text{nr} \ll \gamma_\text{rad}$ (left column of Fig.~\ref{fig:dipoles}), we find aligning dipoles parallel to the transport direction leads to significant improvements in robustness but only moderate improvements in peak transport efficiency at large $J_b$. Although the \cell{prism} geometry was extremely robust against disorder at large $J_b$ in the previous section (when $\gamma_\text{nr}=0$), the inclusion of even a relatively small non-radiative loss rate (e.g. $\gamma_\text{nr} = 0.1\gamma_\text{rad}$) leads to a large drop in robustness, with many of the perturbed networks performing significantly worse than the single chain. This occurs because the many of the perturbed on-site energy configurations generate significant transport bottlenecks via their effect on the phonon-mediated relaxation rates. In the previous ($\gamma_\text{nr}=0$) case, these bottlenecks were irrelevant since transport was occurring through the aforementioned `dark chain' which protected the system from almost all (radiative) losses, even when transport was particularly slow. 

In contrast, the dipole-dependent \dcell{prism} version demonstrates far greater robustness, owing to stronger inter-unit-cell coupling (due to Eq.~\eqref{eq:dipole-coupling}) which leads to increased spatial delocalization of the system eigenstates and, therefore, faster and more uniform eigenstate transition rates throughout the chain.

In the second regime, where $\gamma_\text{nr} = \gamma_\text{rad}$, we find that the $U_\text{prism}$ geometry performs far worse (on average) than the simpler $U_\text{mono}$ case, but, once again, this performance drop off due to non-radiative losses can be somewhat mitigated by including intrinsic dipole effects (middle column of Fig.~\ref{fig:dipoles}). The peak around $J_b = 4$ in the bottom middle panel of Fig.~\ref{fig:dipoles} occurs due to the competition between maximizing dark state populations during transport (to mitigate radiative loss) and avoiding any transport bottlenecks (to mitigate non-radiative loss) -- see Appendix~\ref{apdx:bottlenecks} for details.

Finally, (and unsurprisingly) once $\gamma_\text{nr} \gg \gamma_\text{rad}$ we find that the dipole-independent systems (both $U_\text{mono}$ \& $U_\text{prism}$) perform significantly worse than any other case discussed thus far. However, these large $\gamma_\text{nr}$ cases also exhibit the greatest improvement in average transport performance due to the addition of site dipoles (right column of Fig.~\ref{fig:dipoles}); further highlighting the fact that dipole effects are able to partially protect against non-radiative loss (again, due to the previously mentioned increase in eigenstate delocalization and therefore faster phonon-mediated transport). In Appendix~\ref{apdx:coupling-disorder} we show that adding moderate amounts of random disorder to the dipole orientations can often lead to further enhancements in transport efficiency in the $\gamma_\text{nr} \gg \gamma_\text{rad}$ regime when compared with the perfectly aligned dipole configuration.

\section{Conclusion \& Outlook}

In this work we have demonstrated that it is possible to effectively eliminate radiative recombination from exciton transport networks through careful choice of the repeated unit-cell geometry within the system. By tuning only the intra-unit-cell coupling ($J_b$), the eigenstates of the system were shown to form two distinct, energetically-separated subsets; with almost all of the optical dissipation occurring in the upper band of eigenstates -- thereby allowing long-range transport to proceed via the low energy `dark chain'. This phenomenon was also found to be highly robust to relatively large perturbations in the energetic structure of the system Hamiltonian; enabling simultaneously efficient \textit{and} robust quantum transport.

Furthermore, by including the effects of intrinsic site dipoles and non-radiative recombination, we have shown that aligning site dipoles along the direction of transport can result in excitons traversing the system more quickly, thereby speeding up transport and helping to mitigate the detrimental effects of non-radiative losses. This finding has the potential to appreciably improve the performance of photovoltaic devices, where non-radiative losses significantly inhibit exciton diffusion lengths and therefore degrade energy efficiency~\cite{QT:OPV-review, QT:OPV-nr-loss-1, QT:perovskite-solar-cell-nr-loss}.

While we have focused on well-justified weak coupling approaches for simulating quantum dynamics in this work, it is worth noting that studies involving more rigorous simulation methods -- such as numerically exact tensor network approaches~\cite{OQS:PollockCausalTensorNetworkPI, OQS:multi-bath-TEMPO, OQS:SomozaDAMPF} (which have recently been applied to chain-like systems~\cite{OQS:Fux-chain-thermalization-TEMPO}) or other non-Markovian techniques~\cite{OQS:DeVegaReview} -- may lead to further interesting insights into the interplay between coherent and dissipative dynamics within these quantum transport systems. On the other hand, in the $J_b \gg J_a$ limit, a course-graining-based analysis (similar to that found in~\cite{Renorm-LHC-chain}) may provide a simpler model describing the effects of both unit cell geometry and intrinsic disorder within these transport systems. By mapping the system Hamiltonian used in this work onto a quantum well superlattice model~\cite{Intro:Superlattice-textbook, Intro:Superlattice-textbook-2}, analytic expressions for the system's eigen-structure may also be obtainable.

The generic open quantum systems model developed here, as well as the relative simplicity of the dark state protection mechanism, leads to a transport network which is amenable to experimental implementation in a variety of physical systems. 
As a concrete example, the engineered molecular chain designs in~\cite{QT:molecular-energy-grad} could be realised with inter-chain couplings of $10-300\units{meV}$ and the energy gradient could be tuned such that $\delta E \sim J_a$ via an applied electric field. Similarly, in polar TMO-based materials such as $LaVO_3/SrVO_3$ heterostructures, it has been estimated that $J_a \approx 200\units{meV}$ and $\delta E \approx 3 J_a$~\cite{QT:Electric-field-TMOs_2}, though this ratio can likely be tuned such that $\delta E \sim J_a$ via the application of an external bias.
This simplicity also opens the door to the prospect of combining the findings presented here with other proposals for maximizing transport efficiency~\cite{SD-JCP-biased-chains} or achieving long-range transport~\cite{QT:Plenio-PRX} -- possibly leading to even greater enhancements in both transport efficiency and robustness.

Finally, we anticipate that the underlying physical features elucidated in this work will be of relevance to practical technologies, such as organic photovoltaics, molecular electronics, or other nanoscale systems in which lossless transfer of energy -- or, equivalently, classical information -- is essential.

\begin{acknowledgements}
We thank Dominic Branford for useful discussions. This work was supported by EPSRC grants no. EP/L015110/1 and EP/T007214/1. The open source QuantumOptics.jl package was used for numerical simulations~\cite{QO.jl}.
\end{acknowledgements}

\begin{center}
\rule{0.95\columnwidth}{0.5pt}
\rule{0.95\columnwidth}{0.5pt}
\rule{0.95\columnwidth}{0.5pt}
\end{center}

\appendix

\section{Transport Model Details}

\subsection{Dimensionless Rescaling}
\label{apdx:dimensionless-model}

As stated in the main text, we formulate our transport model such that it is dimensionless, and therefore generally applicable to a variety of physical systems. Specifically, we rescale all model parameters to express them relative to the inter-unit cell coupling $J_a$. Since all unit-cells are equally spaced within our model, in practical terms we can simply set $J_a = 1$ and reinterpret the physical units of all other model parameters as being `per unit inter-chain coupling'. 

To illustrate this more explicitly, we can write out the \textit{non-rescaled} Hamiltonian as 
\begin{equation}
    \label{eq:H-sys-primed}
    \begin{split}
        \hat{H}'_{s} = &\sum_{\mu=1}^N \big[ (N-\mu) \delta E' + E_0' \big] \outerproduct{\mu, i}{\mu, i} + E_g' \outerproduct{g}{g} \\ 
        &+ \sum_{\mu \neq \nu, i, j} \frac{J_a'}{|r'_{\mu, i} - r'_{\nu, j}|^3} \outerproduct{\mu, i}{\nu, j} \\
        &+ \sum_{\mu, i \neq j} \frac{J_b'}{|r'_{\mu, i} - r'_{\mu, j}|^3} \outerproduct{\mu, i}{\mu, j} ~,
    \end{split}
\end{equation}
where all symbols have the same meaning as Eq.~\eqref{eq:H-sys} in the main text apart from that the primed quantities here denote those with inherent dimension. If we then divide Eq.~\eqref{eq:H-sys-primed} through by $J_a'$, and set the nearest neighbour inter-unit-cell separation to 1 distance unit, we end up with a dimensionless system Hamiltonian with appropriately redefined parameters. For example, by setting the (dimensionless) on-site energy gradient to be $\delta E = 1$ in the main text, we actually mean that it is equal in magnitude to the coherent coupling between nearest-neighbours in adjacent unit cells which are separated by unit distance. All other energy parameters, as well as the \textit{intra}-unit-cell coupling $J_b$ referenced in the main text, are interpreted in the same way.

This interpretation also applies to the various system-environment interaction rates (e.g. $\gamma_\text{phonon}, \gamma_\text{inj}, \gamma_\text{ext}, \gamma_\text{rad}$ \& $\gamma_\text{nr}$); however, the phonon bath temperature $T_{ph}$ requires slightly more careful consideration. In this case we work in units of $k_b = 1$ which allows us express the Bose-Einstein distribution as $n_{BE}(\omega, T) = (\exp[\omega/T_{ph}] - 1)^{-1}$ and therefore interpret the dimensionless $T_{ph}$ as `the ratio of the thermal energy scale of the vibrational environment to the inter-unit-cell nearest neighbour coupling strength'. (Similarly, $\omega$ is the ratio of energy eigenstate detuning to $J_a$.)

Another effect of this dimensionless reparametrization is that it will modify the units of times and steady state currents in our model. However, since we are only concerned with steady state quantities, and focus on comparing the current flowing through the various multi-site geometries \textit{relative to the single site case}, this is not relevant to our work.

\subsection{Parameter Values}
\label{apdx:param-vals}

The default parameter values used throughout this work (unless otherwise stated in individual figures or captions) are shown in Table~\ref{table:param-values}.

\begin{table}
    \begin{tabularx}{0.45\textwidth}{
        | >{\centering\arraybackslash}X 
        | >{\centering\arraybackslash}X |
        }

        \hline
        \textbf{\textit{Parameter}} & \textbf{\textit{Value}} \\
        \hline
        $\delta E$ & 1                  \\
        $\ ^\dagger E_0$ & 100          \\
        $E_g$ & 0                       \\
        $\gamma_\text{rad}$ & 0.01      \\
        $\gamma_\text{nr}$ & 0          \\
        $\gamma_\text{phonon}$ & 0.01   \\
        $T_\text{ph}$ & 2.5875          \\
        $\Gamma$ & 0.4                  \\
        $\gamma_\text{inj}$ & $10^{-6}$ \\
        $\gamma_\text{ext}$ & 0.021     \\
        \hline
    \end{tabularx}
    \caption{
    Default parameter values used for the dimensionless model described in Sec.~\ref{sec:model}. $\ ^\dagger$See Appendix~\ref{apdx:lindblad-processes} for justification. 
    }
    \label{table:param-values}
\end{table}
\hfill \\
The parameter value of $T_{ph}$ is chosen such that it physically corresponds to a temperature of $300 \units{K}$ at $J_a = 10 \units{meV}$. The injection and extraction rates ($\gamma_\text{inj}$ \& $\gamma_\text{ext}$) are chosen such that the single excitation approximation remains justified across all of our results (i.e. by explicitly checking that the ground state population $P_g > 0.95$ in all numerical calculations). The absolute magnitudes of the exciton currents in all of our results could be trivially scaled upwards or downwards by adjusting these two rates, but this would adversely affect the validity of our underlying physical assumptions.

\subsection{Dimensionless Dipole-dependent Couplings}

The simple interpretation of the coherent couplings in the dimensionless parametrization of our model is less clear cut once intrinsic dipole moments are introduced in Sec.~\ref{sec:dipoles}. In this case, we simply choose to keep both $J_a = 1$ and the inter-unit-cell separation at unity. This means that the actual coherent couplings between identical sites in neighbouring units cells may be different from unity due to the alternative form of $V_{ij}$ for coupled dipoles [due to Eq.~\eqref{eq:dipole-coupling}].

\section{Effective Lindblad Processes}
\label{apdx:lindblad-processes}

As mentioned in Sec.~\ref{sec:model}, we make use of a number of phenomenological, Lindblad-like system-environment interactions (denoted using the calligraphic $\hat{\mathcal{A}}$) in the construction of our transport model. In practical terms, these are constructed using the appropriate Hermitian interaction operator $\hat{A}$, combined with a spectral density of the form 
\begin{equation}
    \label{eq:Sw-flat-down}
    S_\alpha (\omega) = \gamma_\alpha \Theta(\omega) ~,
\end{equation}
where $\alpha$ denotes the specific system-env interaction and $\Theta$ is the Heaviside step function. By ensuring that all singly-excited eigenstates of the system have an energy greater than the ground state energy $E_g$ -- which is achieved in practice by ensuring $E_0$ is sufficiently large in Eq.~\eqref{eq:H-sys} (we are free to do this since the only energy dependence in Eq.~\eqref{eq:Sw-flat-down} is within $\Theta$) -- this form of $S_\alpha$ converts two-way Hermitian operators linking the ground and excited manifolds into simpler unidirectional operators.

As a concrete example, take the extraction process operator from the $i^{th}$ site of the $N^{th}$ unit cell, which is implemented as 
\begin{align}
    \hat{A}_\text{ext} &= \outerproduct{N, i}{g} + \outerproduct{g}{N, i} \\ \nonumber \\
    S(\omega) &= \gamma_\text{ext} \Theta(\omega) ;
\end{align}
leading to the one-way extraction process given in Eq.~\eqref{eq:A-ext}. For processes which are one-way in the opposite direction (e.g. the injection process in our model), we simply use $S_\alpha (\omega) = \gamma_\alpha (1 - \Theta(\omega))$ in place of Eq.~\eqref{eq:Sw-flat-down}.

\section{Bloch-Redfield Master Equation}
\label{apdx:BRME}

In Fig.~\ref{fig:multi-chain} we include some steady state current data calculated using a non-secular Bloch-Redfield Master Equation (BRME), in order to verify that our Pauli Master Equation model is accurate. The BRME is a microscopically derived, weak-coupling master equation (see Sec.~3.3 of~\cite{OQS:BreuerPetruccione}) which, in our implementation, takes the form
\begin{equation}
    \partial_t \rho(t) = -i [\hat{H}_s, \hat{\rho}] + \hat{R} \rho,
\end{equation}
where the Bloch-Redfield tensor $\hat{R}$ is given by
\begin{equation}
    \hat{R} = \sum_{\alpha, \omega, \omega'} S_\alpha(\omega) \big( A_\alpha(\omega) \rho A^\dagger_\alpha(\omega') + A^\dagger_\alpha(\omega') A_\alpha(\omega) \rho \big) + H.c. ~,
\end{equation}
with the operators $\hat{A}$ and spectral densities $S(\omega)$ having the same meaning as Eq.~\eqref{eq:W-matrix}. We use the numerical implementation provided in the QuantumOptics.jl package~\cite{QO.jl}.

\section{Phonon Rates}
\label{apdx:phonon-rates}

\begin{figure*}
    \includegraphics[scale=0.72]{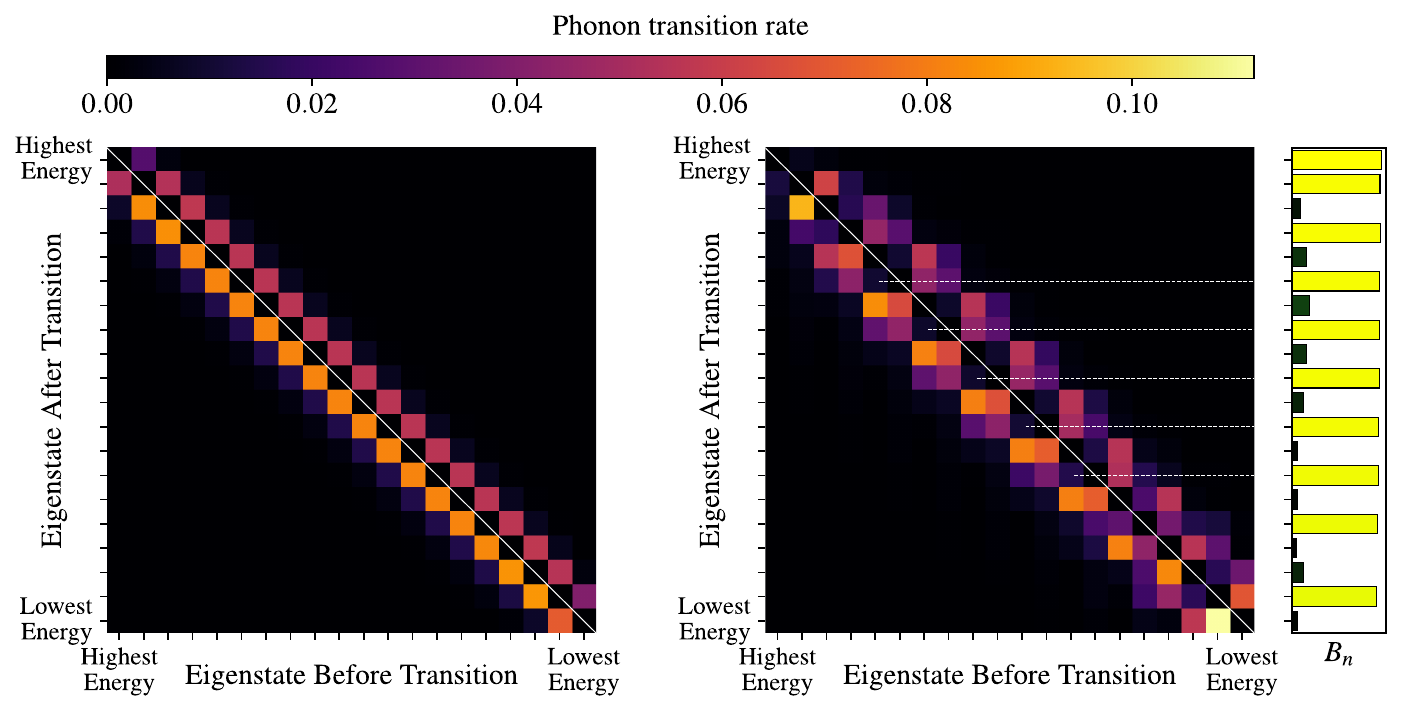}
    \caption{
        Phonon-mediated eigenstate transition rates for (\textit{left}) a single chain system \cell{mono} consisting of 20 unit cells and  (\textit{right}) a \cell{dimer} system consisting of 10 unit cells where certain nearest-neighbour transitions are suppressed. The suppressed transitions correspond to those which would otherwise populate bright states (as shown by the dashed horizontal lines connecting to the brightness bar plot on the right) leading to improved dark state protection and transport performance compared with the single chain case.
    }
    \label{fig:phonon-rates}
\end{figure*}

As shown in the main text, even the simplest \cell{dimer} case with $J_b = J_a$ leads to improved transport performance over the \cell{mono} case. This is due to changes in the phonon-mediated eigenstate transition rates between bright and dark states within the system, with the intra-unit-cell dimerization in the \cell{dimer} case leading to a suppression of phonon processes which would otherwise increase bright state population and therefore reduce transport efficiency. A direct comparison of \cell{mono} and \cell{dimer} phonon rates [i.e. the phonon-related contributions to the $W_{mn}$ matrix of Eq.~\eqref{eq:W-matrix}] is shown in Fig.~\ref{fig:phonon-rates}.

\section{Triangular Prism Eigenstate Structure}
\label{apdx:prism-eigenstructure}

\begin{figure}
    \includegraphics[scale=0.56]{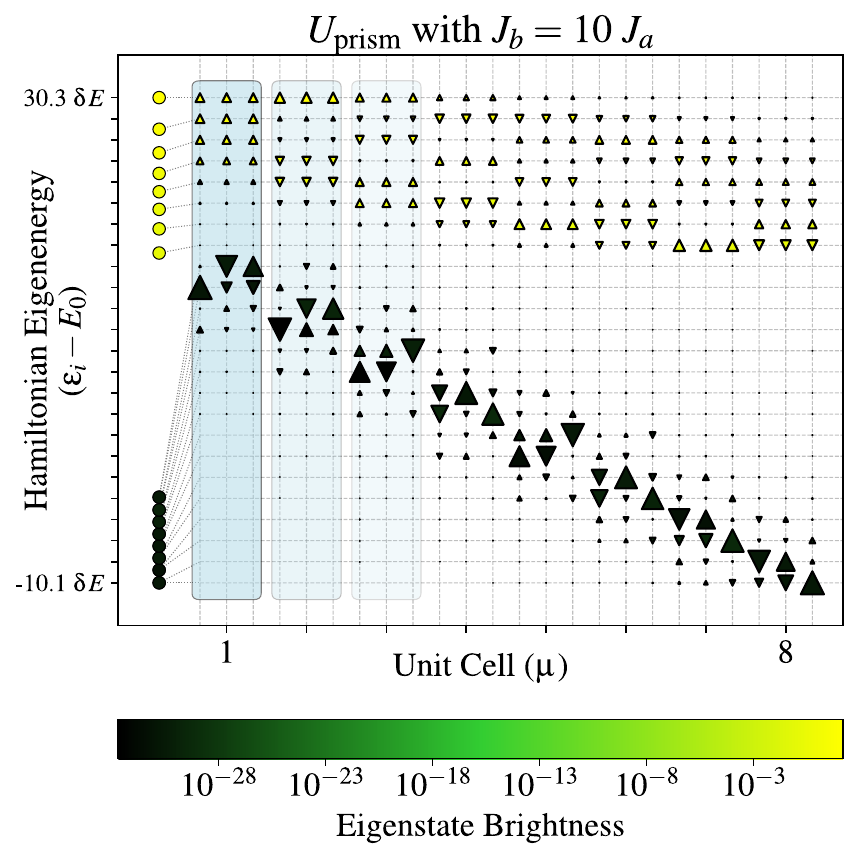}
    \caption{
        Eigenstate structure of the \cell{prism} system demonstrating the energetic separation of eigenstates into bright and dark chains once $J_b \gg J_a$. See caption of Fig.~\ref{fig:eigenbasis} for explanation of plot format.
    }
    \label{fig:prism-eigenstructure}
\end{figure}

Fig.~\ref{fig:prism-eigenstructure} illustrates that the \cell{prism} unit cell geometry leads to a similar eigenstate structure to the \cell{dimer} case shown in Fig.~\ref{fig:eigenbasis}. When $J_b \gg J_a$, the `dark chain' in the lower half of the energy spectrum leads to the effectively lossless transport which is the main result of this work.

\section{Dimer vs Prism Robustness}
\label{apdx:prism-robustness}

\begin{figure}
    \includegraphics[scale=0.66]{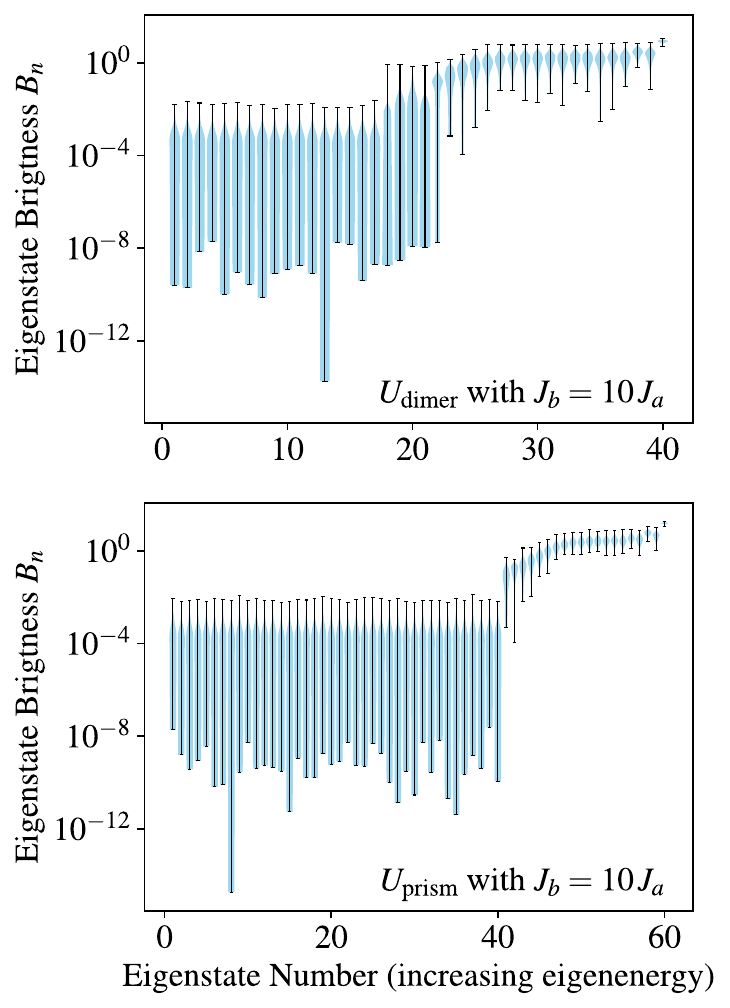}
    \caption{
        Comparison of the eigenstate brightness robustness to on-site energy perturbations between the dimer and prism unit cell geometries (for systems with $N=20$ units cells each). The \cell{dimer} geometry has approximately the same number of bright/dark states, whereas the \cell{prism} case has roughly twice as many dark as bright states. Results were calculated with $10^3$ random realizations of disorder with $\sigma = 0.9 \delta E$ [c.f.~Eq.~\eqref{eq:H-dis}].
    }
    \label{fig:brightness-robustness}
\end{figure}

As shown in bottom row of Fig.~\ref{fig:multi-chain}, the \cell{prism} geometry exhibits far greater robustness to on-site energy perturbations compared with the \cell{dimer} case, even though they both give rise to a dark chain of low energy eigenstates when $J_b \gg J_a$. The origin of this improved robustness for the prism can be explained by considering Fig.~\ref{fig:brightness-robustness}, which shows the distribution of each eigenstate's brightness under the effect of on-site energy perturbations. The most notable difference is in the relative number of bright and dark states in each system, with the prism case containing approximately twice as many dark states as bright. The other notable difference is within the distributions of the highest energy dark states (i.e. those closest in energy to the bright chain) where we see that, in the dimer case, there is an increased tendency to mix with the bright states when the on-site energies are perturbed. This, in turn, can be explained by observing that the prism geometry leads to a larger energetic separation between dark and bright chains compared with the dimer geometry at the same $J_b$ value (as seen in the bottom panels of Fig.~\ref{fig:eigenbasis}). Therefore, larger on-site energy perturbations would be required in the prism case to observe the same level of detrimental eigenstate mixing.

\section{Dipole-dependent Models}

\subsection{Transport Bottlenecks}
\label{apdx:bottlenecks}

\begin{figure}
    \includegraphics[scale=0.58]{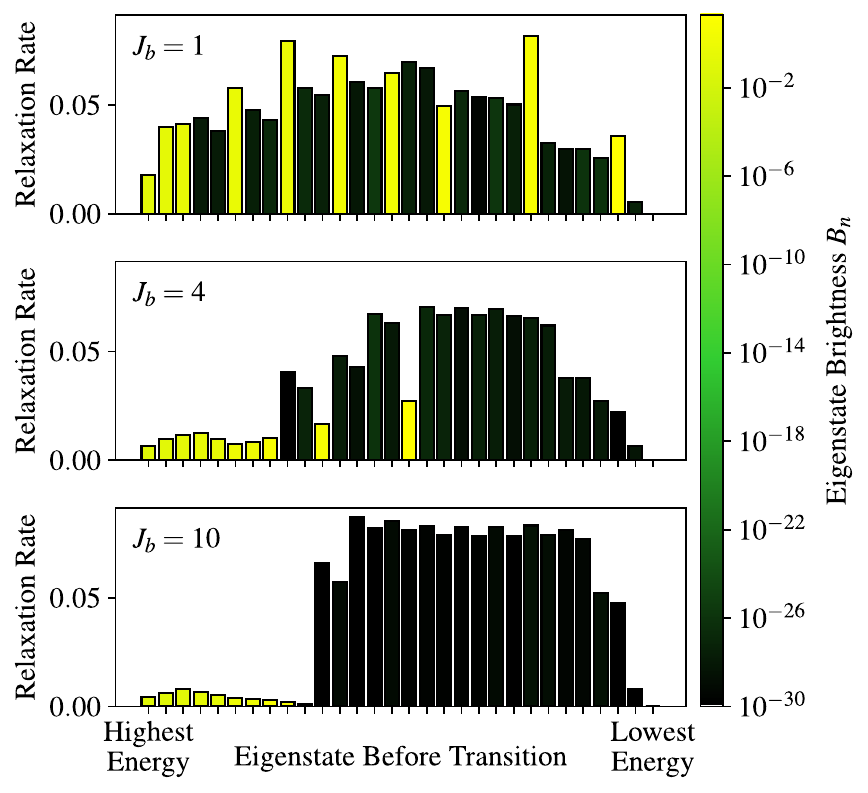}
    \caption{
        Illustration of the phonon-mediated relaxation rates (i.e. sum over all downward energy transitions) from each system eigenstate in a \cell{prism} geometry at three different $J_b$ values. The $J_b = 10$ panel has too many bottlenecks (specifically in transitioning from the bright chain to the dark chain) which increases non-radiative recombination and therefore explains the transport efficiency peak around $J_b = 4$ in Fig.~\ref{fig:dipoles}d.
    }
    \label{fig:SM-bottlenecks}
\end{figure}

As explained in the main text, the primary consideration for avoiding the detrimental effects of non-radiative loss within our transport model is to avoid creating any transport `bottlenecks' (i.e.~places where phonon-mediated eigenstate transitions are particularly slow). The two main factors which affect these eigenstate transition rates are the spatial overlap and energy separation between the two eigenstates in question. Therefore, the fact that increasing $J_b$ in our model leads to a large energetic separation between bright and dark subsections of the Hilbert space (as described in the discussion surrounding Fig.~\ref{fig:eigenbasis}) can serve as a illustrative example of how these bottlenecks can arise.

This effect is illustrated in Fig.~\ref{fig:SM-bottlenecks}, where we see a clear bottleneck developing between bright and dark states when $J_b = 10$. This data is generated using the same transport model as the bottom middle panel of Fig.~\ref{fig:dipoles}; and therefore clearly demonstrates that the peak transport performance around $J_b = 4$ in that scenario is a result of the trade-off between low-energy dark chain transport versus a uniformisation of the phonon-mediated relaxation transition rates which facilitate quick transport.

Another example of the competition between dark state protection and transport bottlenecks can be seen in the \dcell{prism} with $\gamma_\text{nr} = 10 \gamma_\text{rad}$ (bottom right panel of Fig.~\ref{fig:dipoles}) where, upon close inspection, we can see that many of the perturbed energy configurations actually perform better than the clean, unperturbed system -- an effect which is not observed at lower $\gamma_\text{nr}$ values. By examining the effects of on-site energy perturbations on the phonon-mediated relaxation rates from each eigenstate, we find that the additional energetic disorder tends to create a more uniform distribution of relaxation rates and often significantly increases the slowest rates in particular. This is clearly beneficial when non-radiative loss is dominant over radiative loss since the slowest rates are exactly where the previously discussed detrimental bottlenecks occur. Therefore, by alleviating these bottlenecks, the perturbed configurations will (on average) exhibit better steady state currents.

\subsection{Other Unit Cell Geometries}
\label{apdx:sheets-and-dipoles}

\begin{figure*}
    \includegraphics[scale=0.63]{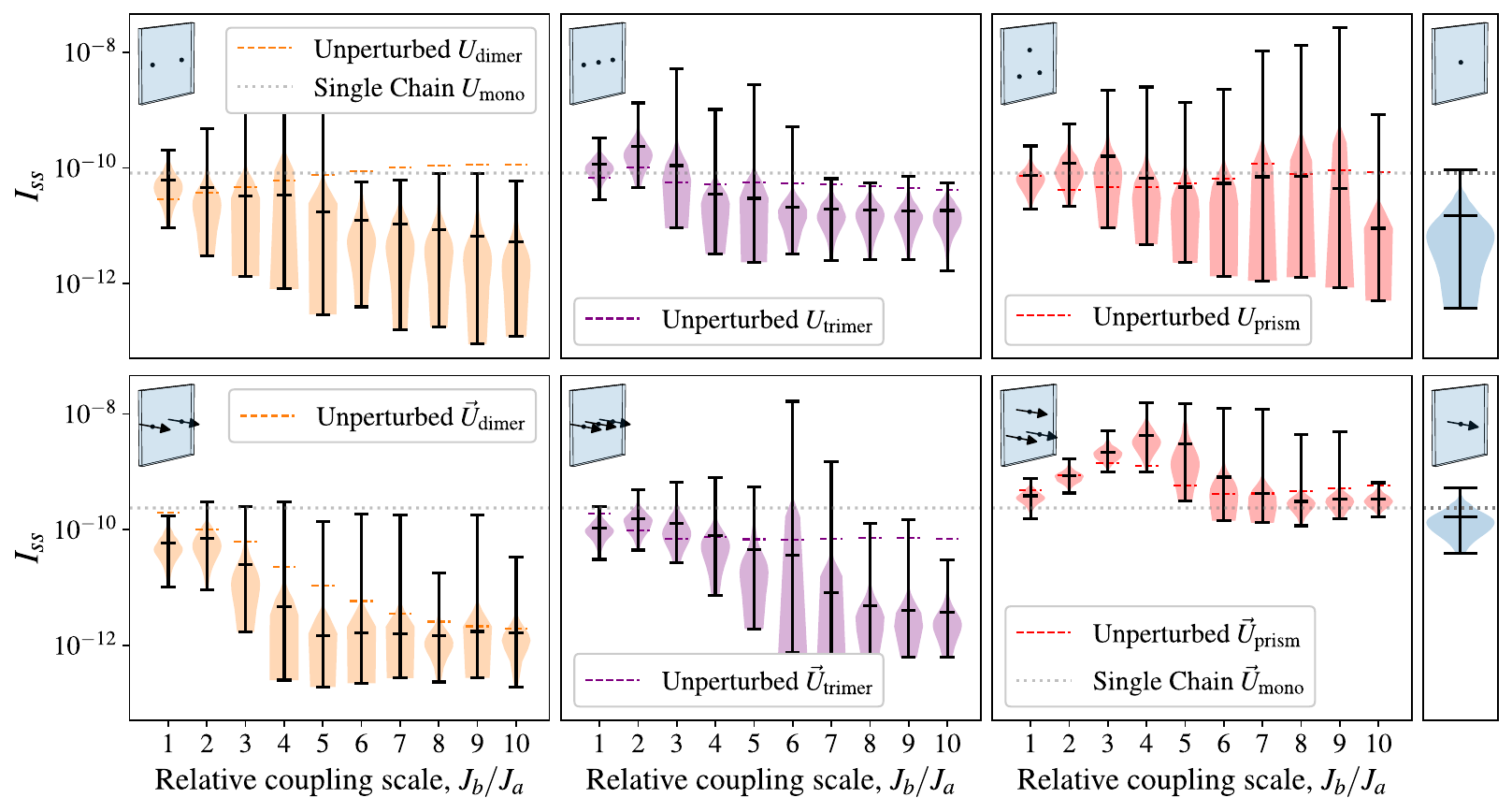}
    \caption{
        Transport efficiency and robustness to on-site energy perturbations as a function of $J_b$ with the inclusion of non-radiative loss processes. The inclusion of dipole-dependent effects is (on average) detrimental to the transport efficiency of the \cell{dimer} (left column) and \cell{trimer} (middle column) geometries, in contrast to the positive effects on the \cell{prism} case (right column -- same parameters as Fig.~\ref{fig:dipoles} b \& d but included here for ease of comparison) which were discussed in detail in Sec.~\ref{sec:dipoles}. Furthest right-hand column shows the effects of energy perturbations and site dipoles on the single chain \cell{mono} system. Non-radiative rates are set at $\gamma_\text{nr} = \gamma_\text{rad}$ for all panels.
    }
    \label{fig:SM-dipoles}
\end{figure*}

Sec.~\ref{sec:dipoles} of the main text revealed that by aligning the intrinsic dipole moments of all sites parallel to the direction of transport in the \cell{prism} geometry, some of the detrimental effects of non-radiative recombination processes could be mitigated. In Fig.~\ref{fig:SM-dipoles} we demonstrate that the same dipole alignment does not have the same beneficial effects on the \cell{dimer} and \cell{trimer} unit cell geometries.

We also find that other simple dipole configurations (such as aligning all dipoles parallel to each other but \textit{orthogonal} to the transport direction) do not lead to consistent improvements in transport performance. The reasons for this generally depend on the specific details of the system -- such as the employed excitation scheme and the relative orientations of unit cells and dipole moments -- however, as mentioned in the main text a more detailed analysis or explicit optimization of the dipole orientations may lead to further interesting systems which are well protected from both radiative and non-radiative losses.

\section{Eigenbasis Injection \& Extraction}
\label{apdx:eigen-inj}

\begin{figure}
    \includegraphics[scale=0.65]{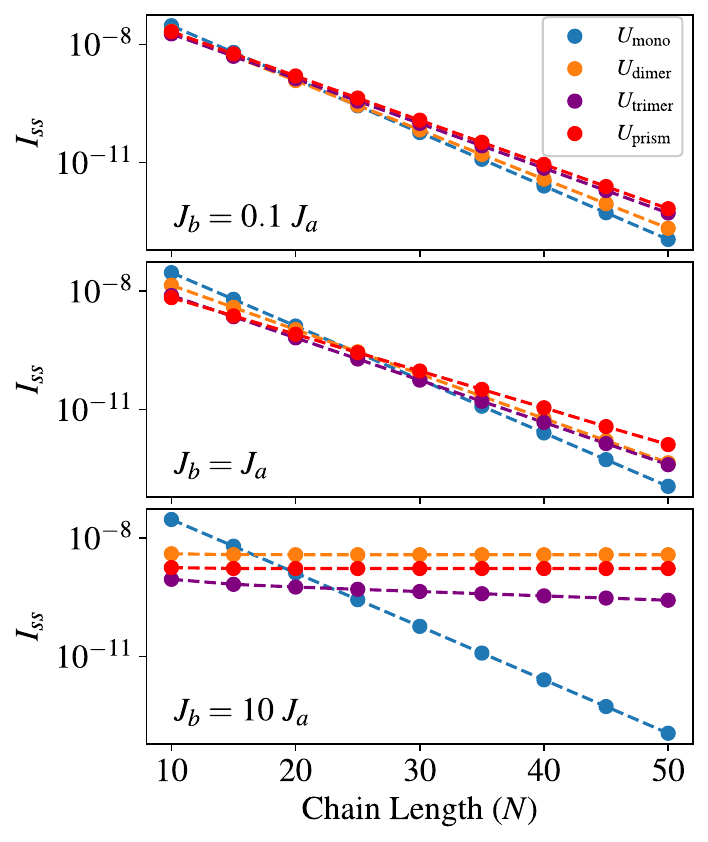}
    \caption{
        Steady state current vs chain length for each unit-cell geometry with injection and extraction processes occurring in the energy eigenbasis. The key qualitative features are unchanged from Fig.~\ref{fig:multi-chain} in the main text, with transport efficiency becoming effectively independent of chain length at large values of $J_b$.
    }
    \label{fig:eigen-inj}
\end{figure}

As discussed in the main text, our transport model uses phenomenological site-basis operators (i.e.~near-field F\"orster coupling) to facilitate the injection and extraction of excitons from the system. In some physical systems, such as those which capture and transport solar photons, a more accurate description of the injection process would instead populate some mixture of high-energy bright states (e.g.~a Gaussian weighted mixture centered near the top of the chain). 

To verify that this alternative description does not invalidate our results, in this section we modify our injection and extraction processes to instead operate in the energy eigenbasis. Specifically, we inject excitations into the highest energy (bright) state of the system, which is localized near unit cell $\mu = 1$, and extract from the lowest energy eigenstate (localized near $\mu = N$). 

Fig.~\ref{fig:eigen-inj} shows the steady state current vs chain length data generated using this alternative injection and extraction model and confirms that, in the $J_b \gg J_a$ regime, both the \cell{dimer} and \cell{prism} cases still exhibit transport efficiency which is effectively independent of chain length.

The most notable difference between Fig.~\ref{fig:eigen-inj} and the equivalent main text result (Fig.~\ref{fig:multi-chain}) is that the single chain (\cell{mono}) system performs better than any of the multi-site unit cells at short chain lengths. This can be explained by considering that exciting the highest energy eigenstate will lead to non-negligible initial population on sites 2, 3 \& 4 (see Fig.~\ref{fig:single-chain}) which, for short chains, is advantageous since it means that the initial injection process has already `transported' the excitation along a considerable fraction of the chain. For longer chains, this shortcut is no longer viable since it skip over a smaller fraction of the total transport distance, therefore the `dark chain' transport mechanism elucidated in the main text will perform more favourably for long-range transport.

\section{Robustness to phonon parameter variations}
\label{apdx:phonon-param-disorder}

\begin{figure}
    \centering
    \includegraphics[scale=0.5]{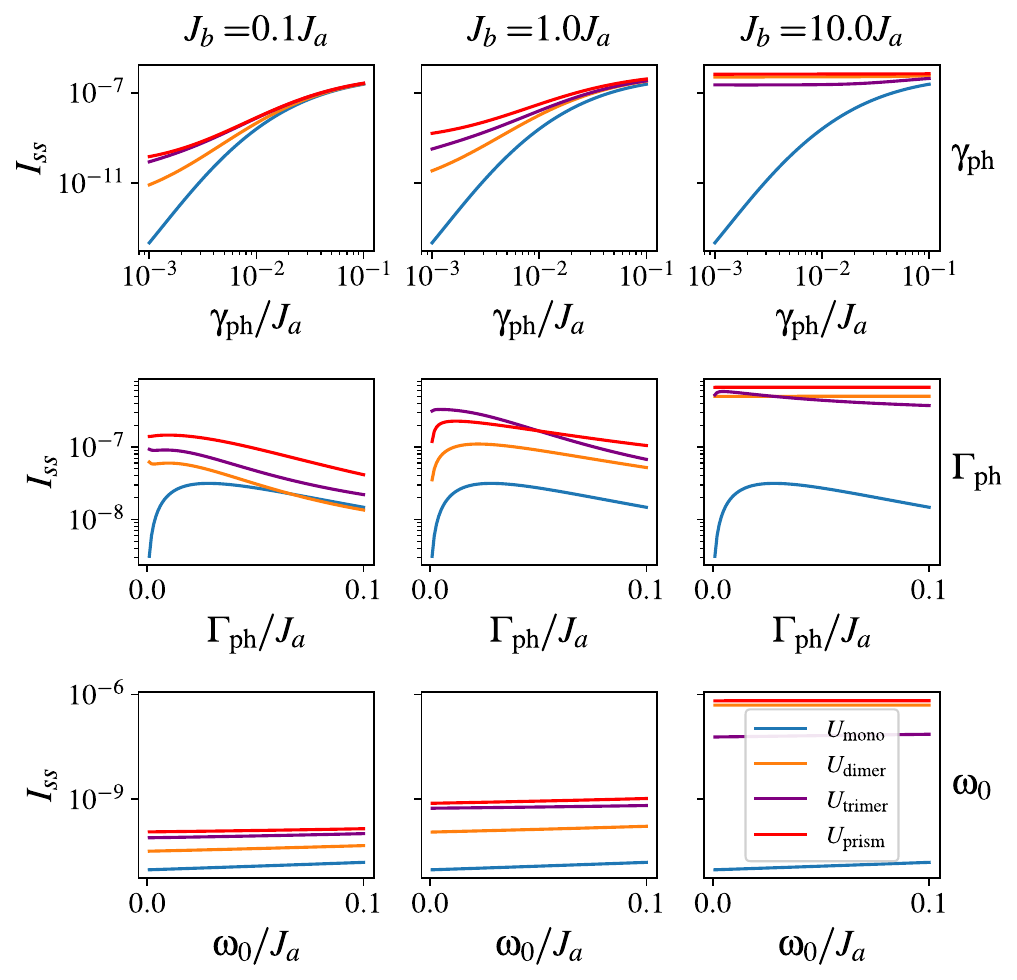}
    \caption{
        Steady state current as a function of the three key phonon spectrum parameters defined in Eq.~\eqref{eq:Sw-DL} with each column corresponding to a different $J_b$ coupling regime and each row to variations in a different phonon parameter. Across all panels the \cell{prism} structure still leads to better steady state currents than the other unit cell geometries in almost all cases.
    }
    \label{fig:phonon-params}
\end{figure}

In real systems, the precise parameter values governing the vibrational bath with which our transport system interacts are unlikely to be known exactly. Therefore, in this section we show that the main conclusion of this work, namely that multi-site unit cells exhibit far better transport performance than the single-site chain, holds while sweeping across a range of values for the phonon coupling strength $\gamma_\text{ph}$, the spectrum width $\Gamma_\text{ph}$ and the phenomenological cutoff frequency $\omega_0$ within the Drude-Lorentz spectral density of Eq.~\eqref{eq:Sw-DL}. 

Firstly, varying $\gamma_\text{ph}$ will merely re-scale the rates of all phonon-mediated eigenstate transitions. Therefore, when these transitions constitute a non-negligible bottleneck in the transport process, increasing the coupling $\gamma_\text{ph}$ should lead to a commensurate increase in the steady state current -- this effect can be seen in the top row of Fig.~\ref{fig:phonon-params}. In the $J_b \gg J_a$ regime, the efficiency of all three multi-site unit cell geometries is largely unaffected by changes in the phonon coupling strength, since this regime corresponds to the `dark-chain transport' scenario discussed in the main text where there is negligible radiative loss and `slow' phonon mediated transport can still be highly efficient.

The second row of Fig.~\ref{fig:phonon-params} shows the steady state current variations as a function of phonon spectrum width $\Gamma_\text{ph}$. The peak in $I_{ss}$ around $\Gamma_\text{ph} = 0.03 J_a$ can be understood by considering the fact that increasing this parameter will lower the magnitude of the phonon spectrum peak while simultaneously increasing the spectrum's width. A lower peak will, as discussed above, lead to slower phonon transition and can therefore lead to poorer transport performance if detrimental loss processes are present. However, a wider spectrum will mean that transitions between non-neighbouring (in energy) states can happen more readily; this can be beneficial in overcoming specific bottlenecks (such as transitioning quickly from the high-energy `bright chain' into the lower energy `dark chain'). Therefore, a trade-off will exist and a specific value of $\Gamma_\text{ph}$ will be optimal. Despite this trade-off, Fig.~\ref{fig:phonon-params} shows that the multi-site unit cell geometries are advantageous in most regimes.

Finally, the bottom row of Fig.~\ref{fig:phonon-params} shows that the system is largely robust to changes in the phenomenological cutoff frequency $\omega_0$. As before, we see that the multi-site unit cells consistently give rise to better transport performance, particularly in the $J_b \gg J_a$ regime.

\section{Robustness to coherent coupling disorder}
\label{apdx:coupling-disorder}

\begin{figure}
    \centering
    \includegraphics[scale=0.7]{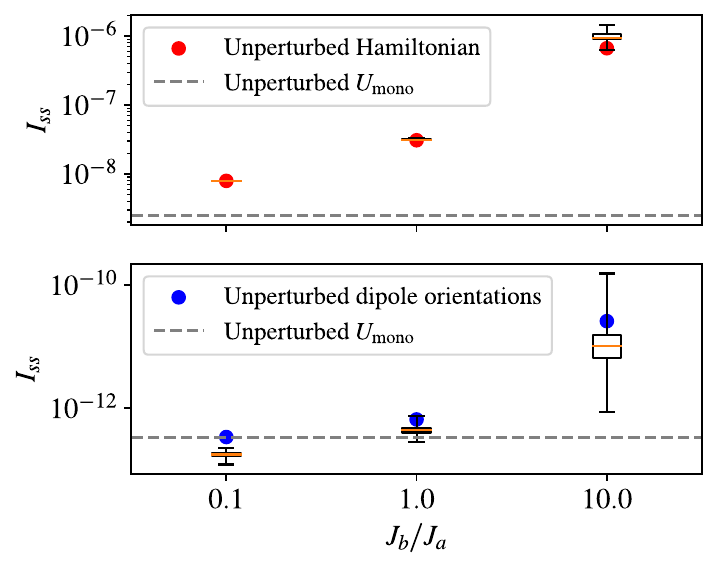}
    \caption{
        Robustness of optimal \cell{prism} geometry to perturbations in the coherent couplings within the system Hamiltonian (1000 realisations of random disorder). In the top panel, couplings are perturbed directly at the level of the individual Hamiltonian elements which leads to very little variation in steady state current. This panel uses a transport model with only radiative losses present (i.e. $\gamma_\text{nr} = 0$). In the bottom panel we set $\gamma_\text{nr} = 0.1$ and look at disorder in site dipole orientations. In this scenario random disorder leads to far greater variation in steady state current; however, the \cell{prism} geometry still performs far better than the \cell{mono} system when $J_b \gg J_a$.
    }
    \label{fig:H-param-disorder}
\end{figure}

As well as robustness to the on-site energies within the Hamiltonian as discussed in Sec.~\ref{sec:robust-to-Es}, we can also show that the best performing \cell{prism} geometry is robust to perturbations in the off-diagonal elements of the Hamiltonian.

The simplest way to do so is to perturb each off-diagonal element of the Hamiltonian with some random disorder sampled from a Gaussian distribution. We take a distribution with standard deviation given by 10\% of the magnitude of the corresponding Hamiltonian element. In other words, each Hamiltonian element $H_{ij}$ ($i > j$) is modified as
\begin{equation} 
    \tilde{H}_{ij} = H_{ij} + \Delta(0, \frac{H_{ij}}{10})
\end{equation}
where $\Delta(0, \sigma)$ is sampled from a Gaussian distribution with 0 mean and a standard deviation of $\sigma = 0.1 H_{ij}$. We then mirror this disorder along the leading diagonal in order to preserve the Hermiticity of the perturbed Hamiltonian. 

By applying this procedure with 1000 random realizations of off-diagonal disorder, we obtain the data shown in the upper panel of Fig.~\ref{fig:H-param-disorder}. This clearly illustrates that the \cell{prism} geometry is robust to arbitrary disorder in the off-diagonal coupling elements of the Hamiltonian.

While the above prescription illustrates that the transport performance is robust to arbitrary coupling disorder, we now turn to another type of disorder which involves perturbations to the individual site dipole orientations. The bottom panel of Fig.~\ref{fig:H-param-disorder} shows the distribution of steady state currents at three different $J_b$ values when the site dipole orientations are subject to small random perturbations in the $\hat{y}$ and $\hat z$ directions (while still being predominantly aligned along the $\hat{x}$ direction). Specifically, the perturbed dipole moments are given by
\begin{equation}
    \vec{d}_i = \frac{1}{\mathcal{N}} \left( \hat{x} + \Delta_1(0, 0.2)\hat{y} + \Delta_2(0, 0.2)\hat{z} \right)
\end{equation}
where $\Delta_{1, 2}$ are random variables drawn from a Gaussian distribution with zero mean and standard deviation $\sigma = 0.2$ and the normalization constant $\mathcal{N} = \sqrt{1 + \Delta_1^2 + \Delta_2^2}$ ensure the dipoles still have unit magnitude. When perturbing the Hamiltonian couplings in this way, the variations in steady state current are significantly larger (as seen in the bottom panel of Fig.~\ref{fig:H-param-disorder}) but despite this, the disordered prism geometry still leads to far better transport performance than \cell{mono} in the $J_b \gg J_a$ regime.


%

\end{document}